\documentclass[fleqn,usenatbib]{mnras}

\usepackage{mathptmx}
\usepackage[T1]{fontenc}

\usepackage{graphicx}	% Including figure files
\usepackage{amsmath}	% Advanced maths commands
\usepackage{amssymb}	% Extra maths symbols

\usepackage{graphicx}
\usepackage{tabularx, array}
\usepackage{adjustbox}
\usepackage[outdir=./]{epstopdf}

\usepackage{float}
\usepackage{txfonts}
\usepackage{natbib}
\usepackage{longtable}

\usepackage{rotating}
\usepackage{lscape}
\usepackage{appendix}
\usepackage{txfonts}
\usepackage{xcolor}
\usepackage{booktabs}
\def\apj{ApJ}
\def\aj{AJ}
\def\apjl{ApJL}
\def\apjs{ApJS}
\def\mnras{MNRAS}

\def\aap{A\&A}

\def\pasp{PASP}

\def\MgFe{[${\rm MgFe}$]$'$}
\def\kms{$\rm km\;s^{-1}$}
\def\Fem{$\langle$Fe$\rangle$}
\def\Alm{$\langle$Al$\rangle$}

\title[NIR spectroscopic indices]{Near-infrared spectroscopic indices for
  unresolved stellar populations. II. Index measurements}

\author[D. Gasparri et al.]{
D. Gasparri,$^{1}$\thanks{E-mail: daniele.gasparri@postgrados.uda.cl}
L. Morelli,$^{1}$
V.~D. Ivanov,$^{2,3}$
P. Fran\c{c}ois,$^{4}$
A. Pizzella,$^{5,6}$
L. Coccato,$^{2}$
E.~M. Corsini,$^{5,6}$
\newauthor
E. Dalla Bont\`a,$^{5,6}$
L. Costantin,$^{7,8}$
and
M. Cesetti$^{9}$
\\
$^{1}$Instituto de Astronom\'{\i}a y Ciencias Planetarias,
  Universidad de Atacama, Copayapu 485, Copiapo\'o, Chile\\
$^{2}$European Southern Observatory, Karl-Schwarzschild-Strasse 2,
  D-85748 Garching bei M\"unchen, Germany\\
$^{3}$European Southern Observatory, Avenida Alonso de C\'ordova 3107,
  Vitacura, Santiago, Chile\\
$^{4}$GEPI, Observatoire de Paris, PSL Research University, CNRS,
  Universit\'e Paris Diderot, Sorbonne Paris Cit\'e, 61 Avenue de l'Observatoire,
  F-75014 Paris, France\\
$^{5}$Dipartimento di Fisica e Astronomia ``G. Galilei'', Universit\`{a} di Padova,
  vicolo dell'Osservatorio 3, I-35122 Padova, Italy\\
$^{6}$INAF-Osservatorio Astronomico di Padova, vicolo dell'Osservatorio 5,
  I-35122 Padova, Italy\\
$^{7}$Centro de Astrobiolog\'ia (CSIC-INTA), Ctra de Ajalvir km 4,
  Torrej\'on de Ardoz, E-28850 Madrid, Spain\\
$^{8}$INAF-Osservatorio Astronomico di Brera, via Brera 28, I-20121 Milano, Italy\\
$^{9}$Independent Scholar, viale degli Aironi 4, I-30021, Caorle (VE), Italy
}

\date{Accepted 2021 August 17. Received 2021 July 22; in original form 2021 April 10.}

\pubyear{2021}

\begin{document}
\label{firstpage}
\pagerange{\pageref{firstpage}--\pageref{lastpage}}
\maketitle

% Abstract of the paper
\begin{abstract}
We measured the equivalent width of a large set of near-infrared (NIR, 0.8--2.4$ \ \mu$m) line-strength indices in the XShooter medium-resolution spectra of the
central regions of 14 galaxies. We found that two aluminum indices Al
at 1.31 $\mu$m and Al1 at 1.67 $\mu$m and the two CO indices CO1 at
1.56 $\mu$m and CO4 at 1.64 $\mu$m are tightly correlated with the velocity dispersion. Moreover, the NIR Al and CO1 indices show strong correlations with the optical Mg2 and Mgb indices, which are
usually adopted as $alpha$/Fe-enhancement diagnostics. The molecular
FeH1 index at 1.58 $\mu$m tightly correlates with the optical \Fem\/
and [MgFe]' indices, which are used as total metallicity diagnostics. The NIR
Pa$\beta$ index at 1.28 $\mu$m has a behaviour similar to the
optical H$\beta$ index, which is a diagnostic of mean age. We defined
two new composite indices, \Alm\/ and [AlFeH], as possible candidates
to be used as NIR diagnostics of total metallicity and $\alpha$/Fe
enhancement. The NIR \Alm\/ index has a strong correlation with the
optical Mg2 and Mgb indices, while the [AlFeH] index is tightly
correlated with the optical \Fem\/ and \MgFe\/ indices.  The
distribution of the data points in the NIR Pa$\beta$-\Alm\/ and
Pa$\beta$-[AlFeH] diagrams mimic that in the optical \MgFe-H$\beta$
and the Mgb-\Fem\/ diagrams, which are widely used to constraint the properties of the unresolved stellar populations. We concluded that
some NIR line-strength indices could be useful in studying stellar
populations as well as in fine-tuning stellar population models.
\end{abstract}

% Select between one and six entries from the list of approved keywords.
% Don't make up new ones.

%%%%%%%%%%%%%%%%%%%%%%%%%%%%%%%%%%%%%%%%%%%%%%%%%%

%%%%%%%%%%%%%%%%% BODY OF PAPER %%%%%%%%%%%%%%%%%%
\section{Introduction}

Since the 1970s when the modern era of galaxy spectroscopy began
  with the introduction of image intensifiers (e.g., Image Dissector
  Scanner at Lick Observatory \citep{robinson72}, many advances were done in the knowledge of
unresolved stellar populations by investigating optical spectra of
galaxies. Nowadays, the line-strength index analysis
\citep[e.g.,][]{worthey1994, morelli2008, vazdekis2010, costantin2019}
and full spectral fitting \citep[e.g.][]{sarzi2006, koleva2009,
  morelli2015} are standard tools to recover the star-formation
history of galaxies.

Although some pioneeristic works on the strongest near-infrared (NIR,
0.8--2.4 $\mu$m) spectral absorption features date back to the 1990s
\citep{silva1994, origlia1997}, only recently the increasing size and
efficiency of NIR detectors have allowed a comprehensive spectroscopic
study of the $IYJHK$ bands making the NIR domain complementary to the
optical range in studying the stellar populations of galaxies.
The need of observing the optical range at high redshifts triggered
the building of the new generation of NIR-optimised spectrograph
\citep[e.g,][]{mobasher2010, cuby2010, cirasuolo2011}, which also
offer the chance of having high-quality data to investigate in detail
the NIR spectral energy distribution of nearby galaxies.

This gives some advantages. For example, the luminosity fraction of
stars of early spectral types diminishes as the wavelength increases
\citep{bica1988} and we can actually isolate the asymptotic giant branch
(AGB) and red giant branch (RGB) components. Indeed, the contribution
of stars outside the AGB-RGB branches is negligible in the $K$ band
for all the galaxies independently of their Hubble type
\citep{kotilainen2012}. Nevertheless, the AGB and RGB phases are still
poorly understood and very difficult to model in the NIR
\citep{rock2017, riffel2019}.
On the other hand, the reduced effect of reddening with respect to the
optical range is a crucial advantage of NIR and it allows to peer into
on highly obscured galaxies, otherwise impossible to be investigated
\citep{engelbracht1998, ivanov2000}.  In this context, analysing NIR
spectra could help to better understand the connection between the presence
of dust and young stellar populations, as found for example by
\citet{peletier2007} in the central regions of early-type
spirals.
The major observational drawback for NIR spectroscopy is the strong
contamination produced by telluric absorption and atmospheric emission
lines which overlap to interesting spectral features of galaxies even
at low redshift \citep{francois2019}.

Different sets of NIR line-strength indices were developed since
  the early 1980s \citep{jones1984, bica1987, cenarro2003} mainly
  focusing on the calcium triplet (CaT) and hydrogen Paschen lines
  \citep[see][for an extensive review]{cenarro2001}.

\citet{mannucci2001} investigated some NIR features, at low resolution
($300<R<600$) and facing severe line blending, in a sample of 28 nearby
galaxies covering all the Hubble sequence. \citet{silva2008} studied
some $K$ band line-strength indices, including the strong CO molecular
feature at 2.30 $\mu$m and Ca and Na indices, in a sample of 11
early-type galaxies in the Fornax cluster. \citet{cesetti2009}
performed low-resolution ($R = 1000$) spectroscopy of 14 early-type
galaxies in the wavelength range 1.5--2.4 $\mu$m, covering the strong
Mg feature at 1.50 $\mu$m. \citet{kotilainen2012} focused onto the
strongest Mg, Si, and CO features detected in the central regions of a
sample of 29 quiescent spiral galaxies, which they measured at a
resolution of $R \sim 600$ in the $H$ and $K$ bands.
\citet{rock2015b} defined and investigated some new line-strength
indices in the $JHK$ bands, including those of the aluminium 1.30
$\mu$m line and other Mg lines. \citet{riffel2019} measured several
line-strength indices from $I$ to $K$ band in 16 luminous infrared
spiral galaxies observed at a resolution of $R= 1000$ and in 19
early-type galaxies from literature.

With the increasing number and quality of NIR spectra, large stellar
libraries have been assembled, which either limited to the $JHK$ bands
\citep{ivanov2004} or cover the entire NIR range, like the NASA Infrared
Telescope Facility (IRTF) spectral library \citep{rayner2009,
  villaume2017} and the more recent XShooter spectral library
\citep[XSL][]{arentsen2019}.
These are the base for defining and calibrating new NIR line-strength
indices and for building more complete and reliable single stellar
population (SSP) models. The end products of the SSP models are
  spectra of single burst of star formations at fixed age,
  metallicity, and abundance of $\alpha$-elements \citep{rock2015,
    rock2016, vazdekis2016}. If the assumption of a SSP or of
  dominating SSP is not valid (e.g., for E+A galaxies or for
  particular galaxy components) the SSP models can be combined with
  different luminosity of mass weights to mimic the spectra of
  multi-component stellar populations \citep[e.g.,][]{capems04} and to
  also account for an additional gaseous component
  \citep[e.g.,][]{sarzetal06}. This can be done using either the
  typical line-strength indices \citep{mehlert2003,sanchez2006}, or
  combining the analysis of a set of line-strength indices
  \citep[e.g.,][]{zibetti2017,costantin2019}, or through a full spectral fitting analysis
  \citep[e.g.][]{morelli13, costantin2021}.

This is relatively new field of extragalactic research and it still
lacks a reliable system of line-strength indices to be used as a
diagnostic tools like the Lick/IDS ones \citep{faber1985, worthey1994,
  worthey1997, thomas2003}.  With the aim of filling this void, in
\cite{cesetti2013} and \cite{morelli2020} we developed a set of
spectroscopic diagnostics for stellar physical parameters based on NIR
spectral features in the wavelength range 0.8--5 $\mu$m.

In \cite[][Paper I hereafter]{francois2019}, we presented a sample of
high signal-to-noise ratio (SNR) galaxy spectra obtained at medium
resolution ($R\sim 4000-5000$) with the XShooter spectrograph
\citep{guinard2006} mounted at the Very Large Telescope
(VLT) of the European Southern Observatory (ESO). These spectra
simultaneously map the optical and NIR ranges of galaxy spectral
energy distribution, making easier the direct comparison of spectral
properties of unresolved stellar populations in these regimes. The
purpose of this paper is to analyse these XShooter spectra focusing
onto the NIR bands to investigate the wide set of line-strength
indices defined by \citet{cesetti2013} and \citet{morelli2020}.

We structured the paper as follows. We introduce our dataset in
Section \ref{sec:data}. We describe the measurements of the NIR
line-strength indices in Section \ref{sec:measure}. We investigate
our final set of 40 NIR line-strength indices in Section
\ref{sec:results} and show the correlation among the NIR line-strength
indices and with central velocity dispersion of the galaxies.  We
focus on the age and metallicity indicators in Section
\ref{sec:correlation_optical} and show the correlations with with the
optical Lick/IDS line-strength indices. Finally, we discuss the
results in Section \ref{sec:discussion} and present our conclusions in
Section \ref{sec:conclusions}.

\section{Spectroscopic data}
\label{sec:data}

Medium-resolution spectroscopy was performed performed with the UVB
($R \sim 4000$), VIS ($R \sim 5400$) and NIR ($R \sim 4300$) arms of
the ESO XShooter spectrograph for a sample of 14 nearby galaxies
(Prop. Id. 086.B-0900, PI: Cesetti, M.). The morphological type of the
sample galaxies ranges from E to Sc, central stellar velocity
dispersion is between 36 and 335 \kms, and distance is comprised
between 13 and 62 Mpc.

The spectra were obtained along the major axis of the galaxies and were
co-added along the spatial direction to map a central region of $1.5
\times 1.5$ arcsec$^2$, which corresponds to an area ranging from $65
\times 65$ pc$^2$ to $430 \times 430$ pc$^2$ depending on distance,
with a typical SNR$\,\sim\,100$ \AA$^{-1}$. In Paper I, we measured
the Mg, Fe, and H$\beta$ line-strength indices of the Lick/IDS system
\citep{faber1985, worthey1994}, $\langle$Fe$\rangle$ mean index
\citep{gorgas1990}, \MgFe\/ combined index \citep{thomas2003}, and
their uncertainties following \cite{morelli2004, morelli2012,
  morelli2016}. We derived the stellar population properties of the
sample galaxies using the SSP models by \cite{johansson2010} and found
that ages range from 0.8 to 15 Gyr and metallicities ([Z/H]) from $-0.39$ to
$+0.55$ dex.

Considering the SNR, spatial and spectral resolution, wavelength range
of the spectra as well as the Hubble type, age, and metallicity of the
sample galaxies, our data have an unprecedented quality with respect
to previous works and they represent an excellent resource for
studying stellar populations in the NIR bands.

\section{Definition and measurement of NIR line-strength indices}
\label{sec:measure}

\subsection{Line-strength index measurement}

In the past 15 years, different authors defined and tested an
increasing set of NIR line-strength indices as possible diagnostic
tools \citep[e.g. ][and references therein]{silva2008, cesetti2009,
  rock2015, riffel2019}.  Here, we adopted the definitions given by
\citet{cesetti2013} for the line-strength indices in $I$ and $K$ bands
and by \citet{morelli2020} for the $Y$, $I$, and $H$ bands, which rely
onto the concept of sensitivity maps to identify the spectral features
which are more sensitive to age and metallicity. Considering the
spectral range of our data, we identified and measured 75 NIR
line-strength indices. This is one of the largest and most complete
set of NIR line-strength indices investigated so far.

For each sample galaxy, we measured the equivalent width (EW) of all
the above NIR line-strength indices as done by \citet{cesetti2013} and
\citet{morelli2020} for the IRTF stellar library. In this paper,
  we adopted the same spatial aperture used for the measuring the
  optical line-strength indices in Paper I.
We measured the SNR of each NIR line-strength index in the two
adjacent continuum bands. Ten out of 14 galaxies have a SNR$\,>\,100$
\AA$^{-1}$ for all the line-strength indices, whereas the
line-strength indices of NGC~3423, NGC~4415 and NGC~7424 have
SNR$\,<\,$30 SNR \AA$^{-1}$. In the case of NGC 1600, the
contamination due to the residuals of the sky subtraction prevented us
to measure all the NIR line-strength indices.
The errors on indices were derived from photon
statistics and CCD read-out noise, and calibrated by means of Monte
Carlo simulations. For each line-strength index in each spectrum, we
generated 1000 simulations.

We corrected the EWs to zero velocity dispersion following the method
of \citet{silva2008} and \citet{cesetti2009}.  We broadened the
spectra of the giant stars with a spectral type ranging from K0 to M3
in the IRTF library up to 400 \kms\ with bins of 50 \kms.  For each
NIR line-strength index, we calculated the correction coefficients
with a spline interpolation of the average broadened EW. In addition,
as a sanity check we considered different samples of stars including
the supergiants stars and/or extending to late G-type stars, and
considering stellar spectra from the XSL library. The correction
coefficients are consistent within the uncertainties and match those
calculated by \cite{cesetti2013} and \citet{morelli2020} for giant
K-type stars.
The uncertainties on the correction coefficients were estimated
following a similar approach of the coefficient determination. We
calculated the rms of the EWs of the stellar spectra with respect to
the mean values for any velocity-dispersion broadening value of any
NIR line-strength index. We made a spline interpolation of the rms
values as a function of broadening and considered this function as
representative of the uncertainties associated to the correction
coefficients relative to the considered line-strength index.
  The uncertainties in the index EW increase according to the applied
  multiplicative factor as pointed out by \cite{trager2000}.  For the
total EW uncertainties, we also considered the uncertainties
associated to the correction coefficients, which we added in
quadrature to the intrinsic errors.

As the result of the velocity dispersion correction, we identified
three groups of NIR line-strength indices:

\begin{itemize}
 \item The Ca2, Nadk, and CO12 indices are almost insensitive to the
   velocity dispersion broadening with a maximum variation with
   respect to the zero velocity-dispersion value smaller than $20\%$
   at $\sigma = 400$ \kms\ \citep{cesetti2013}.  The rms of the
   correction coefficients is significantly small ($< 10\%$).

\item The Mg, Al, Si, and molecular line-strength indices, with the
  exception of few TO line-strength indices, are very sensitive to the
  velocity dispersion broadening with a maximum variation with respect
  to the zero velocity-dispersion value larger than $50\%$ at $\sigma
  = 400$ \kms\/ \citep{morelli2020} although the rms of the correction
  coefficients is small ($< 20\%$).

\item Many Fe and H line-strength indices are very weak and extremely
  sensitive to the velocity dispersion broadening and the rms of the
  correction coefficients is very large ($>50\%$).

\end{itemize}

We decided to investigate only NIR line-strength indices with a
SNR$\,>\,20$ \AA$^{-1}$ following \citet{morelli2020} and to exclude
from further investigation the following indices:

\begin{itemize}

\item Pa2, Pa3, Pa4, FeClTi, CSi, Pa$\epsilon$, TiOA, TiOB, Fe, VO,
  Siy, Pa$\gamma$, C, K2B, K1, Fe2, Br13, FeH2, Br11, FeI, since they
  have correction coefficients for the velocity dispersion broadening
  with an rms$\,>\,50\%$;

\item FeCr, FeI, Br$\delta$, Ca1k, Fe23, Sik, Ca2k, Ca3k, Ca4k, Pa5,
  Pa6, Na, and Br$\gamma$, which could be contaminated by
  telluric absorption and atmospheric emission;

\item Br16 and CO5, because they are clearly contaminated by other
  spectral features.

\end{itemize}

The index denomination follows the definitions by \cite{cesetti2013}
and \cite{morelli2020}, which are based on the main element that
produces the spectral feature. When more than one line-strength index
is identified with the same name, we added a suffix that refers to the
NIR band where it is observed. We did not detect NIR emission lines in
any of the spectra of the sample galaxies.

\begin{table*}
\centering
\caption{The final set 40 NIR line-strength indices investigated in
  this paper. The references are 1 = \citet{morelli2020}; 2 =
  \citet{cenarro2001}; 3 = \citet{cesetti2013}; 4 =
  \citet{ivanov2004}; 5 = \citet{silva2008}; 6 = \citet{cesetti2009};
  7 = \citet{riffel2019}; 8 = \citet{origlia1993}; 9 =
  \citet{conroy2012, villaume2017}; 10 = \citet{mclean2003,
    cushing2005}, 11 = \citet{rock2015}.}
\label{tab:index_def}

\begin{tabular}{@{}c@{\hspace{0.6cm} }c@{\hspace{0.6cm}}c@{\hspace{0.6cm}}c@{\hspace{0.6cm}}c@{\hspace{0.6cm}}c@{\hspace{0.6cm}}c@{}}
\hline
\hline
Index   & Dominated by                          & Line limits   & Blue continuum    &   Red continuum & Main  & Additional\\
        name &                                  & ($\mu$m)         & ($\mu$m) & ($\mu$m)  & Reference & Reference\\
\hline
Pa1     & H{~\sc i} (n=3)            &~ 0.8461--0.8474~&~ 0.8474--0.8484&~  0.8563--0.8577 & 2 &            \\
Ca1     &   Ca{~\sc ii}              &~ 0.8484--0.8513~&~  0.8474--0.8484&~ 0.8563--0.8577 & 2 &         \\
  Ca2     &   Ca{~\sc ii}           &~  0.8522--0.8562~&~ 0.8474--0.8484&~  0.8563--0.8577 & 2 &       \\
 Ca3     &   Ca{~\sc ii}           &~   0.8642--0.8682~&~ 0.8619--0.8642&~  0.8700--0.8725 & 2 &             \\
  Mgi      &   Mg{~\sc i}          &~   0.8802--0.8811~&~ 0.8776--0.8792&~  0.8815--0.8850 & 3 &       \\
Ti     & Ti{~\sc i} &~ 0.9780--0.9795~&~ 0.9750--0.9760&~ 0.9800--0.9810                   & 1 &        \\
FeH     & FeH &~ 0.9900--0.9950~&~ 0.9840--0.9850&~ 0.9985--0.9995                         & 1 & 9,10 \\
Pa$\delta$     & HI &~ 1.0040--1.0067~&~ 1.0020--1.0030&~ 1.0067--1.0077                   & 1 &         \\
FeTi   & Fe{~\sc i}, Ti{~\sc i} &~ 1.0390--1.0408~&~ 1.0198--1.0210&~ 1.0438--1.0446       & 1 &         \\
CN     & CN            &~ 1.0868--1.0882~&~ 1.0640--1.0650&~ 1.0892--1.0902                & 1 &         \\
Sr     & Sr{~\sc ii}   &~ 1.0913--1.0923~&~ 1.0892--1.0902&~ 1.0978--1.0988                & 1 &        \\
K1A     & K{~\sc i}    &~ 1.1670--1.1714~&~ 1.1560--1.1585&~ 1.1716--1.1746                & 1 & 9,10 \\
K1B     & K{~\sc i}    &~ 1.1765--1.1800~&~ 1.1716--1.1746&~ 1.1805--1.1815                & 1 & 9,10 \\
Mgj     &   Mg{~\sc i} &~  1.1820--1.1840~&~ 1.1805--1.1815&~ 1.1855--1.1875               & 1 & 7 \\
Sij     & Si{~\sc i} &~ 1.1977--1.2004~&~ 1.1910--1.1935&~ 1.2050--1.2070                  & 1 &          \\
SiMg     & Si{~\sc i}, Mg{~\sc i} &~ 1.2070--1.2095~&~ 1.2050--1.2070&~ 1.2050--1.2070     & 1 &            \\
K2A     & K{~\sc i} &~ 1.2415--1.2455~&~ 1.2350--1.2380&~ 1.2460--1.2490                   & 1 &  7,10          \\
Pa$\beta$     &   H{~\sc i} &~   1.2795--1.2840~&~ 1.2755--1.2780&~ 1.2855--1.2873         & 1 &               \\
Al     &   Al{~\sc i} &~   1.3115--1.3168~&~ 1.3050--1.3075&~ 1.3230--1.3250               & 1 &  9,11 \\
Mg1h     &   Mg{~\sc i} &~  1.4850--1.4910~&~ 1.4830--1.4850&~ 1.4910--1.5000              & 1 &  7 \\
Mg2h     &   Mg{~\sc i} &~   1.5000--1.5080~&~ 1.4910--1.5000&~ 1.5100--1.5120             & 1 &  7,4 \\
CO1     & $^{12}$CO(2,0) &~ 1.5570--1.5635~&~ 1.5480--1.5500&~ 1.5930--1.5940               & 1 &  7 \\
Br$_{15}$     & H{~\sc i} &~ 1.5670--1.5720~&~ 1.5480--1.5500&~ 1.5930--1.5940              & 1 &     \\
Mg3h     & Mg{~\sc i} &~ 1.5730--1.5800~&~ 1.5480--1.5500&~ 1.5930--1.5940                 & 1 &   7 \\
FeH1     &   FeH &~  1.5820--1.5860~&~ 1.5480--1.5500&~ 1.5930--1.5940                     & 1 &  4,7,8 \\
Sih     &   Si{~\sc i} &~  1.5870--1.5910~&~ 1.5480--1.5500&~ 1.5930--1.5940               & 1 &  4,7,8 \\
CO2     & $^{12}$CO(2,0) &~ 1.5950--1.6000~&~ 1.5930--1.5940&~ 1.6160--1.6180               & 1 &   7 \\
CO3     &   $^{12}$CO(2,0) &~  1.6180--1.6220~&~ 1.6160--1.6180&~ 1.6340--1.6370            & 1 &  8 \\
CO4     &   $^{12}$CO(2,0) &~  1.6390--1.6470~&~ 1.6340--1.6370&~ 1.6585--1.6605            & 1 &          \\
Fe3     & Fe{~\sc i} &~ 1.6510--1.6580~&~ 1.6340--1.6370&~ 1.6585--1.6605                  & 1 &          \\
Al1     &   Al{~\sc i} &~  1.6705--1.6775~&~ 1.6585--1.6605&~ 1.6775--1.6790               & 1 &       \\
COMg   &   $^{12}$CO(2,0), Mg{~\sc i} &~  1.7050--1.7130~&~ 1.6920--1.6960&~ 1.7140--1.7160 & 1 &  4 \\
Br$_{10}$  & H{~\sc i} &~ 1.7350--1.7390~&~ 1.7250--1.7280&~ 1.7440--1.7480                 & 1 &       \\
Mg1k         & Mg{~\sc i} &~ 2.1040--2.1110~&~ 2.1000--2.1040&~ 2.1110--2.1150             & 4 &       \\
Nadk&   Na{~\sc i} &~  2.2000--2.2140~&~ 2.1934--2.1996&~ 2.2150--2.2190                   & 6 &       \\
FeA         & Fe{~\sc i} &~ 2.2250--2.2299~&~ 2.2133--2.2176&~ 2.2437--2.2497              & 5 &   \\
FeB         & Fe{~\sc i} &~ 2.2368--2.2414~&~ 2.2133--2.2176&~ 2.2437--2.2497              & 5 &      \\
Ca$_{\rm d}$ & Ca{~\sc i} &~ 2.2594--2.2700~&~ 2.2516--2.2590&~ 2.2716--2.2888               & 6 &           \\
Mg2k         & Mg{~\sc i} &~ 2.2795--2.2845~&~ 2.2700--2.2720&~ 2.2850--2.2874             & 5 &     \\
CO12   &   $^{12}$CO(2,0) &~  2.2910--2.3070~&~ 2.2516--2.2590&~ 2.2716--2.2888             & 3 &    \\

\hline

\end{tabular}
\end{table*}

 \subsection{Final set of indices}
 \label{sec:set_indices}

The final set of NIR line-strength indices contains 40 entries listed
in Table \ref{tab:index_def}.

We performed a visual inspection of the spectra of the sample galaxies
and our conclusions are in line with previous works
\citep[e.g.][]{silva2008, conroy2012, rock2017, alton2018} and can be
summarised as it follows:

\begin{itemize}

\item As expected \citep{alton2018}, the strongest line-strength
  indices from molecular features are those related to the CO bands,
  in particular the CO12 index at 2.30 $\mu$m. The COMg index at 1.71
  $\mu$m is the strongest non-CO molecular feature. Other molecular
  indices, like Ti, FeTi, CN, and FeH1 are weaker but still
  detectable. The FeH Wing-Ford index, a known gravity diagnostic
  \citep{schiavon1997, conroy2012}, is very weak for almost all the
  sample galaxies confirming the findings by \citet{alton2017}.

\item The strongest line-strength indices from atomic lines are those
  related to Ca, Mg, and Al, in particular the Ca1, Ca2, and Ca2
  indices at $\sim0.85$ $\mu$m, Mg2h index at 1.50 $\mu$m, and Al1
  index at 1.67 $\mu$m. The Nadk index at 2.20
  $\mu$m and Sij index at 1.59 $\mu$m are remarkable features. The Fe
  line-strength indices are generally weak. The line-strength indices
  from atomic lines in the $K$ band are weak, but reside in a spectral
  region free of contamination.

\item Pa1, Pa$\delta$, and Pa$\beta$ are the strongest indices among
  those related to hydrogen. The Pa1 virtually disappears for NGC~1600
  and NGC~3115 due to velocity dispersion broadening, with a
  correction coefficient greater than $80\%$. Therefore, the Pa1 EW
  for these two galaxies has to be considered with caution.  The Br15
  index is the strongest feature of the Brackett series, but it is
  always weaker than the Paschen indices. The Br10 index is very weak
  in all the sample galaxies.

\end{itemize}

The EW and corresponding uncertainty of the NIR line-strength indices
measured with high accuracy for the sample galaxies are listed in
Tables \ref{tab:idx1} (SNR$\,>\,100$ \AA$^{-1}$ ) and \ref{tab:idx2}
(SNR$\,<\,100$ \AA$^{-1}$ ).

\begin{table*}
  \caption{
    The equivalent widths in \AA\ of the NIR line-strength indices of the sample
    galaxies with high SNR ($\,>\,100$ \AA$^{-1}$ ).}
\label{tab:idx1}

\resizebox{\textwidth}{!}{%
\begin{tabular}{cccccccccccc}
\hline
\hline

Index & NGC 584  &   NGC 636  &   NGC 897  &   NGC 1357 &   NGC 1425  &   NGC 1700 &   NC2613  &   NGC 3115 &   NGC 3377 &   NGC 3379 \\
\hline
Pa1	&	0.50	$\pm$	0.12	&	0.56	$\pm$	0.10	&	0.52	$\pm$	0.13	&	0.45	$\pm$	0.07	&	0.35	$\pm$	0.05	&	0.19	$\pm$	0.06	&	0.52	$\pm$	0.09	&	1.04	$\pm$	0.25	&	0.44	$\pm$	0.07	&	0.50	$\pm$	0.12	\\
Ca1	&	1.40	$\pm$	0.12	&	1.49	$\pm$	0.09	&	1.49	$\pm$	0.14	&	1.44	$\pm$	0.07	&	1.46	$\pm$	0.05	&	1.37	$\pm$	0.13	&	1.38	$\pm$	0.08	&	1.23	$\pm$	0.14	&	1.46	$\pm$	0.07	&	1.41	$\pm$	0.13	\\
Ca2	&	3.42	$\pm$	0.08	&	3.61	$\pm$	0.06	&	3.60	$\pm$	0.08	&	3.57	$\pm$	0.05	&	3.67	$\pm$	0.04	&	--	 &	3.43	$\pm$	0.06	&	3.44	$\pm$	0.08	&	3.59	$\pm$	0.05	&	3.31	$\pm$	0.07	\\
Ca3	&	2.46	$\pm$	0.05	&	3.15	$\pm$	0.04	&	2.89	$\pm$	0.06	&	3.12	$\pm$	0.04	&	3.14	$\pm$	0.03	&	--	&	3.15	$\pm$	0.05	&	2.97	$\pm$	0.07	&	3.05	$\pm$	0.04	&	2.87	$\pm$	0.05	\\
Mgi	&	0.36	$\pm$	0.06	&	0.72	$\pm$	0.06	&	0.53	$\pm$	0.07	&	0.73	$\pm$	0.05	&	0.64	$\pm$	0.04	&	1.37	$\pm$	0.13	&	0.72	$\pm$	0.06	&	0.54	$\pm$	0.05	&	0.55	$\pm$	0.04	&	0.88	$\pm$	0.07	\\
Ti	&	0.61	$\pm$	0.04	&	0.53	$\pm$	0.04	&	0.59	$\pm$	0.04	&	0.58	$\pm$	0.03	&	0.62	$\pm$	0.05	&	-0.07	$\pm$	0.03	&	0.75	$\pm$	0.04	&	-0.33	$\pm$	0.20	&	0.26	$\pm$	0.05	&	0.88	$\pm$	0.13	\\
FeH	&	1.56	$\pm$	0.34	&	0.94	$\pm$	0.21	&	2.02	$\pm$	0.45	&	1.04	$\pm$	0.20	&	0.81	$\pm$	0.14	&	3.15	$\pm$	0.75	&	1.01	$\pm$	0.21	&	0.55	$\pm$	0.22	&	0.59	$\pm$	0.12	&	-0.08	$\pm$	0.07	\\
Pa$\delta$	&	0.75	$\pm$	0.11	&	0.63	$\pm$	0.10	&	0.83	$\pm$	0.13	&	0.52	$\pm$	0.07	&	0.82	$\pm$	0.07	&	0.90	$\pm$	0.12	&	0.52	$\pm$	0.12	&	0.84	$\pm$	0.13	&	0.77	$\pm$	0.09	&	0.66	$\pm$	0.12	\\
FeTi	&	0.41	$\pm$	0.04	&	0.40	$\pm$	0.04	&	0.28	$\pm$	0.03	&	0.24	$\pm$	0.03	&	0.22	$\pm$	0.04	&	0.34	$\pm$	0.03	&	0.10	$\pm$	0.02	&	0.37	$\pm$	0.03	&	0.34	$\pm$	0.03	&	0.34	$\pm$	0.03	\\
CN	&	0.62	$\pm$	0.05	&	0.62	$\pm$	0.04	&	0.54	$\pm$	0.05	&	0.43	$\pm$	0.03	&	0.41	$\pm$	0.03	&	0.62	$\pm$	0.05	&	0.50	$\pm$	0.03	&	0.56	$\pm$	0.06	&	0.52	$\pm$	0.03	&	0.59	$\pm$	0.04	\\
Sr	&	-0.14	$\pm$	0.04	&	-0.13	$\pm$	0.03	&	-0.10	$\pm$	0.04	&	-0.23	$\pm$	0.04	&	-0.13	$\pm$	0.02	&	-0.11	$\pm$	0.03	&	-0.03	$\pm$	0.02	&	-0.14	$\pm$	0.04	&	-0.15	$\pm$	0.03	&	-0.14	$\pm$	0.04	\\
K1A	&	0.67	$\pm$	0.06	&	0.48	$\pm$	0.05	&	0.34	$\pm$	0.04	&	0.40	$\pm$	0.04	&	0.79	$\pm$	0.04	&	0.49	$\pm$	0.05	&	0.75	$\pm$	0.04	&	0.46	$\pm$	0.09	&	0.26	$\pm$	0.05	&	0.60	$\pm$	0.09	\\
k1B	&	0.66	$\pm$	0.05	&	-0.28	$\pm$	0.04	&	0.90	$\pm$	0.06	&	0.28	$\pm$	0.04	&	0.20	$\pm$	0.05	&	1.07	$\pm$	0.06	&	0.83	$\pm$	0.05	&	1.01	$\pm$	0.09	&	1.23	$\pm$	0.06	&	0.69	$\pm$	0.05	\\
Mgj	&	1.17	$\pm$	0.07	&	0.28	$\pm$	0.05	&	1.26	$\pm$	0.07	&	0.87	$\pm$	0.07	&	1.72	$\pm$	0.21	&	1.16	$\pm$	0.06	&	0.93	$\pm$	0.08	&	1.12	$\pm$	0.12	&	1.27	$\pm$	0.07	&	1.03	$\pm$	0.08	\\
Sij	&	0.89	$\pm$	0.06	&	1.06	$\pm$	0.06	&	1.09	$\pm$	0.07	&	0.98	$\pm$	0.06	&	0.87	$\pm$	0.04	&	0.89	$\pm$	0.05	&	0.84	$\pm$	0.05	&	0.63	$\pm$	0.06	&	1.02	$\pm$	0.05	&	0.84	$\pm$	0.05	\\
SiMg	&	0.59	$\pm$	0.04	&	0.57	$\pm$	0.04	&	0.51	$\pm$	0.05	&	0.61	$\pm$	0.05	&	0.57	$\pm$	0.03	&	0.45	$\pm$	0.03	&	0.69	$\pm$	0.05	&	0.28	$\pm$	0.05	&	0.46	$\pm$	0.04	&	0.42	$\pm$	0.06	\\
K2A	&	1.06	$\pm$	0.04	&	0.99	$\pm$	0.04	&	1.44	$\pm$	0.06	&	1.02	$\pm$	0.03	&	0.57	$\pm$	0.02	&	0.69	$\pm$	0.03	&	0.81	$\pm$	0.03	&	1.07	$\pm$	0.04	&	0.91	$\pm$	0.02	&	0.97	$\pm$	0.03	\\
Pa$\beta$	&	1.94	$\pm$	0.08	&	1.99	$\pm$	0.08	&	1.52	$\pm$	0.08	&	1.63	$\pm$	0.07	&	1.69	$\pm$	0.05	&	1.80	$\pm$	0.08	&	1.98	$\pm$	0.08	&	1.72	$\pm$	0.08	&	1.67	$\pm$	0.06	&	1.56	$\pm$	0.07	\\
Al	&	1.99	$\pm$	0.08	&	1.54	$\pm$	0.06	&	1.83	$\pm$	0.08	&	1.72	$\pm$	0.06	&	1.04	$\pm$	0.02	&	2.05	$\pm$	0.11	&	1.58	$\pm$	0.05	&	2.04	$\pm$	0.09	&	1.53	$\pm$	0.07	&	2.04	$\pm$	0.09	\\
Mg1h	&	1.48	$\pm$	0.38	&	1.75	$\pm$	0.38	&	1.57	$\pm$	0.43	&	1.73	$\pm$	0.32	&	1.58	$\pm$	0.22	&	1.93	$\pm$	0.53	&	1.28	$\pm$	0.25	&	1.91	$\pm$	0.61	&	1.78	$\pm$	0.37	&	1.72	$\pm$	0.47	\\
Mg2h	&	4.26	$\pm$	0.20	&	4.44	$\pm$	0.16	&	4.34	$\pm$	0.24	&	4.47	$\pm$	0.15	&	3.89	$\pm$	0.10	&	4.39	$\pm$	0.23	&	4.31	$\pm$	0.14	&	4.44	$\pm$	0.32	&	4.45	$\pm$	0.17	&	3.70	$\pm$	0.23	\\
CO1	&	5.02	$\pm$	0.21	&	4.99	$\pm$	0.19	&	5.23	$\pm$	0.23	&	4.38	$\pm$	0.18	&	4.10	$\pm$	0.16	&	4.61	$\pm$	0.20	&	4.32	$\pm$	0.20	&	5.21	$\pm$	0.23	&	4.87	$\pm$	0.19	&	5.14	$\pm$	0.22	\\
Br15	&	1.46	$\pm$	0.28	&	1.96	$\pm$	0.31	&	1.85	$\pm$	0.39	&	1.75	$\pm$	0.26	&	1.38	$\pm$	0.18	&	1.69	$\pm$	0.34	&	1.44	$\pm$	0.25	&	1.24	$\pm$	0.29	&	0.93	$\pm$	0.17	&	1.95	$\pm$	0.39	\\
Mg3h	&	7.20	$\pm$	0.31	&	6.61	$\pm$	0.27	&	6.20	$\pm$	0.29	&	6.68	$\pm$	0.27	&	6.11	$\pm$	0.23	&	6.34	$\pm$	0.28	&	5.76	$\pm$	0.26	&	6.64	$\pm$	0.31	&	5.75	$\pm$	0.24	&	6.40	$\pm$	0.29	\\
FeH1	&	2.64	$\pm$	0.24	&	2.41	$\pm$	0.20	&	2.52	$\pm$	0.26	&	2.47	$\pm$	0.20	&	2.54	$\pm$	0.18	&	2.12	$\pm$	0.21	&	1.94	$\pm$	0.19	&	2.56	$\pm$	0.27	&	2.06	$\pm$	0.18	&	2.56	$\pm$	0.25	\\
Sih	&	4.59	$\pm$	0.16	&	4.15	$\pm$	0.15	&	4.48	$\pm$	0.17	&	4.15	$\pm$	0.15	&	4.48	$\pm$	0.15	&	4.20	$\pm$	0.15	&	4.42	$\pm$	0.17	&	4.39	$\pm$	0.15	&	3.96	$\pm$	0.14	&	4.29	$\pm$	0.16	\\
CO2	&	3.55	$\pm$	0.27	&	3.67	$\pm$	0.24	&	3.34	$\pm$	0.24	&	3.76	$\pm$	0.19	&	3.47	$\pm$	0.16	&	3.56	$\pm$	0.25	&	3.29	$\pm$	0.20	&	3.41	$\pm$	0.29	&	2.71	$\pm$	0.17	&	3.09	$\pm$	0.21	\\
CO3	&	4.85	$\pm$	0.19	&	4.52	$\pm$	0.18	&	4.69	$\pm$	0.18	&	4.94	$\pm$	0.20	&	4.50	$\pm$	0.16	&	4.11	$\pm$	0.18	&	4.27	$\pm$	0.18	&	3.98	$\pm$	0.17	&	4.55	$\pm$	0.17	&	3.74	$\pm$	0.15	\\
CO4	&	3.80	$\pm$	0.08	&	4.27	$\pm$	0.08	&	4.22	$\pm$	0.09	&	3.77	$\pm$	0.09	&	3.94	$\pm$	0.08	&	4.14	$\pm$	0.08	&	3.30	$\pm$	0.07	&	4.31	$\pm$	0.10	&	4.47	$\pm$	0.08	&	4.02	$\pm$	0.08	\\
Fe3	&	-0.20	$\pm$	0.07	&	0.62	$\pm$	0.12	&	0.99	$\pm$	0.23	&	1.17	$\pm$	0.20	&	0.19	$\pm$	0.06	&	0.81	$\pm$	0.19	&	0.35	$\pm$	0.08	&	0.48	$\pm$	0.16	&	1.13	$\pm$	0.18	&	0.21	$\pm$	0.08	\\
Al1	&	4.05	$\pm$	0.38	&	3.88	$\pm$	0.24	&	4.42	$\pm$	0.39	&	3.64	$\pm$	0.17	&	3.39	$\pm$	0.23	&	3.43	$\pm$	0.31	&	3.44	$\pm$	0.21	&	4.45	$\pm$	0.50	&	4.08	$\pm$	0.21	&	4.11	$\pm$	0.41	\\
COMg	&	5.25	$\pm$	0.12	&	5.16	$\pm$	0.11	&	4.71	$\pm$	0.11	&	5.22	$\pm$	0.12	&	5.18	$\pm$	0.10	&	4.59	$\pm$	0.11	&	5.01	$\pm$	0.11	&	4.99	$\pm$	0.12	&	5.02	$\pm$	0.11	&	4.95	$\pm$	0.11	\\
Br10	&	1.24	$\pm$	0.09	&	1.16	$\pm$	0.07	&	1.74	$\pm$	0.11	&	1.25	$\pm$	0.07	&	1.19	$\pm$	0.06	&	1.45	$\pm$	0.09	&	1.08	$\pm$	0.07	&	1.32	$\pm$	0.09	&	1.18	$\pm$	0.08	&	1.13	$\pm$	0.09	\\
MgIk	&	2.02	$\pm$	0.07	&	1.64	$\pm$	0.06	&	1.81	$\pm$	0.07	&	1.77	$\pm$	0.07	&	1.08	$\pm$	0.05	&	1.56	$\pm$	0.05	&	1.97	$\pm$	0.06	&	1.31	$\pm$	0.05	&	1.37	$\pm$	0.05	&	1.20	$\pm$	0.04	\\
Nadk	&	5.22	$\pm$	0.24	&	5.56	$\pm$	0.22	&	4.34	$\pm$	0.23	&	5.28	$\pm$	0.17	&	3.73	$\pm$	0.17	&	5.22	$\pm$	0.25	&	3.89	$\pm$	0.15	&	5.71	$\pm$	0.33	&	5.02	$\pm$	0.19	&	4.12	$\pm$	0.20	\\
FeA	&	1.00	$\pm$	0.06	&	1.49	$\pm$	0.10	&	1.05	$\pm$	0.08	&	1.09	$\pm$	0.08	&	1.27	$\pm$	0.08	&	2.04	$\pm$	0.11	&	1.00	$\pm$	0.07	&	1.18	$\pm$	0.08	&	1.47	$\pm$	0.07	&	1.10	$\pm$	0.08	\\
FeB	&	1.13	$\pm$	0.08	&	1.11	$\pm$	0.10	&	-0.04	$\pm$	0.03	&	0.59	$\pm$	0.07	&	0.60	$\pm$	0.08	&	1.24	$\pm$	0.09	&	0.61	$\pm$	0.06	&	0.30	$\pm$	0.03	&	0.36	$\pm$	0.04	&	0.26	$\pm$	0.04	\\
Cad	&	2.26	$\pm$	0.11	&	1.81	$\pm$	0.17	&	2.62	$\pm$	0.16	&	-0.46	$\pm$	0.20	&	3.60	$\pm$	0.19	&	2.94	$\pm$	0.11	&	1.48	$\pm$	0.13	&	3.57	$\pm$	0.14	&	3.28	$\pm$	0.11	&	2.93	$\pm$	0.12	\\
Mg2k	&	1.15	$\pm$	0.09	&	1.07	$\pm$	0.11	&	0.62	$\pm$	0.07	&	0.96	$\pm$	0.05	&	0.52	$\pm$	0.08	&	0.30	$\pm$	0.05	&	1.52	$\pm$	0.17	&	0.66	$\pm$	0.06	&	0.80	$\pm$	0.05	&	0.67	$\pm$	0.05	\\
CO12	&	23.48	$\pm$	0.21	&	21.47	$\pm$	0.27	&	18.42	$\pm$	0.25	&	24.85	$\pm$	0.33	&	20.40	$\pm$	0.22	&	20.95	$\pm$	0.18	&	23.22	$\pm$	0.23	&	23.40	$\pm$	0.26	&	23.26	$\pm$	0.15	&	22.30	$\pm$	0.20	\\

\hline
\end{tabular}}
\end{table*}

\begin{table}
  \caption{The equivalent widths in \AA\ of the NIR line-strength indices of the
    sample galaxies with low SNR ($\,<\,100$ \AA$^{-1}$ ).}
\label{tab:idx2}

\begin{tabular}{c@{}cccc}
\hline
\hline

Index name &   NGC 1600 &   NGC 3423 &   NGC 4415 &   NGC 7424 \\
\hline
Pa1	&	-0.82 $\pm$ 0.28	&	0.56	$\pm$	0.13	&	0.31	$\pm$	0.05	&	0.25	$\pm$	0.08	\\
Ca1	&	0.98	$\pm$	0.14	&	1.95	$\pm$	0.16	&	1.28	$\pm$	0.06	&	1.35	$\pm$	0.10	\\
Ca2	&	3.23	$\pm$	0.09	&	3.97	$\pm$	0.19	&	3.37	$\pm$	0.07	&	3.51	$\pm$	0.11	\\
Ca3	&	3.07	$\pm$	0.10	&	3.12	$\pm$	0.08	&	3.13	$\pm$	0.09	&	3.84	$\pm$	0.14	\\
Mgi	&	0.57	$\pm$	0.08	&	0.28	$\pm$	0.04	&	0.71	$\pm$	0.05	&	0.48	$\pm$	0.05	\\
Ti	&	0.43	$\pm$	0.07	&	1.02 $\pm$   0.14	&	0.41	$\pm$	0.15	&	0.35	$\pm$	0.08	\\
FeH	&	--	&	--	&	--	&	--	\\
Pa$\delta$	&	--	&	--	&	--	&	--	\\
FeTi	&	0.18	$\pm$	0.05	&	--	&	--	&	--	\\
CN	&	--	&	1.16	$\pm$	0.10	&	0.16	$\pm$	0.10	&	--	\\
Sr	&	--	&	0.05	$\pm$	0.10	&	-0.18	$\pm$	0.06	&	0.14	$\pm$	0.11	\\
K1A	&	--	&	1.16	$\pm$	0.20	&	1.11	$\pm$	0.14	&	1.02	$\pm$	0.33	\\
k1B	&	--	&	0.30	$\pm$	0.20	&	0.61	$\pm$	0.12	&	1.73	$\pm$	0.34	\\
Mgj	&	1.46	$\pm$	0.15	&	1.04	$\pm$	0.11	&	--	&	--\\
Sij	&	1.15	$\pm$	0.12	&	0.24	$\pm$	0.15	&	0.53	$\pm$	0.14	&	0.95	$\pm$	0.23	\\
SiMg	&	--	&	--	&	--	&	--	\\
K2A	&	--	&	--	&	--	&	--	\\
Pa$\beta$	&	1.45  $\pm$   0.10	&	1.97	$\pm$	0.29	&	--	&	3.03   $\pm$   0.50	\\
Al	&	1.99	$\pm$	0.12	&	1.72	$\pm$	0.16 	&	0.86	$\pm$	0.14	&	-0.10	$\pm$	0.49	\\
Mg1h	&	1.04	$\pm$	0.35	&	2.05	$\pm$	0.48	&	2.85	$\pm$	0.24	&	-- 	\\
Mg2h	&	4.19	$\pm$	0.36	&	--	&	--	&	--	\\
CO1	&	--	&	--	&	--	&	--	\\
Br15	&	1.87	$\pm$	0.62	&	1.53   $\pm$   0.51	&	2.38	$\pm$	0.29	&	--	\\
Mg3h	&	5.86	$\pm$	0.38	&	--	&	--	&	--	\\
FeH1	&	2.25	$\pm$	0.38	&	1.52	$\pm$	0.46	&	2.15	$\pm$	0.25	&	0.77	$\pm$	0.78	\\
Sih	&	3.32	$\pm$	0.20	&	3.68	$\pm$	0.45	&	4.17	$\pm$	0.25	&	3.88	$\pm$	0.76	\\
CO2	&	3.44	$\pm$	0.40	&	3.08	$\pm$	0.23	&	--	&	--	\\
CO3	&	4.57	$\pm$	0.23	&	--	&	4.22	$\pm$	0.23	&	5.19	$\pm$	0.43	\\
CO4	&	4.01	$\pm$	0.13	&	1.93	$\pm$	0.20	&	2.96	$\pm$	0.25	&	2.76	$\pm$	0.39	\\
Fe3	&	--	&	--&	1.71	$\pm$	0.29	&	0.02	$\pm$	0.35	\\
Al1	&	4.56	$\pm$	0.60	&	2.28	$\pm$	0.18	&	3.98	$\pm$	0.15	&	1.96	$\pm$	0.26	\\
COMg	&	--	&	3.85	$\pm$	0.19	&	--	&	--	\\
Br10	&	2.29	$\pm$	0.16	&	1.29	$\pm$	0.20	&	0.87    $\pm$	0.20    &	-- \\
MgIk	&	--	&	--	&	--	&	--	\\
Nadk	&	--	&	--	&	--	&	--	\\
FeA	&	--	&	--	&	--	&	--	\\
FeB	&	--	&	--	&	--	&	--	\\
Cad	&	--	&	--	&	--	&	--	\\
Mg2k	&	--	&	--	&	--	&	--	\\
CO12	&	--	&	--	&	--	&	--	\\
\hline
\end{tabular}
\end{table}

\section{Results}
\label{sec:results}

In this section, adopting a completely phenomenological approach, we
present the properties of the NIR line-strength and their correlations
with the velocity dispersion of the galaxies.

\subsection{NIR-NIR correlations}
\label{sec:nirnir}

We started correlating each other the NIR line-strength indices to
identify the most promising diagnostic of the properties of unresolved
stellar populations. We adopted this approach with twofold aim of
investigating the possible linear correlations between the NIR line-strength
indices related to different elements and between the NIR
line-strength indices based on the same element.
We performed a linear regression considering both X
    and Y uncertainties through the data points and we calculated their
Pearson correlation coefficient $R$ with the CORR Pearson function of
the {\tt PANDAS}\footnote{W. McKinney, pandas: a python data
  analysis library, http://pandas.sourceforge.net} Python package.

Fig.~\ref{fig:corr_map_nir} displays the correlation map for the final
set of 40 NIR line-strength indices grouped by elements with
increasing wavelength.  For showing purposes, we cut the colour scale
to $|R|>0.5$. We refer to a {\em moderate\/} correlation if $0.5< |R|<
0.7$ and to a {\em strong\/} correlation when $|R|> 0.70$. The reddest
and bluest squares mark to strongest correlations and
anti-correlations, respectively.

The plot diagonal hosts the correlations of the NIR line-strength
indices with themselves, which give a Pearson coefficient $R = 1$ by
definition.  To avoid spurious correlations due to large EW errors, we
considered only the correlations with $|R| > 0.5$ for NIR
line-strength indices with a SNR$\,>\,20$ \AA$^{-1}$, as done by
  \citet{morelli2020}. In some spectra, some line-strength indices do
  not pass the SNR threshold because they are contaminated by residuals
  of the subtraction telluric absorption and/or atmospheric emission,
  while, in the same spectra, other line-strength indices always have
  SNR$\,>\,20$ \AA$^{-1}$ (e.g. Pa$\beta$). Furthermore, we
considered only the correlations that still hold with $|R| > 0.5$ when
one random galaxy is removed from the analysis to account for the
small number statistics of our galaxy sample.

\begin{figure*}
\centering
\includegraphics[width=18truecm,angle=0]{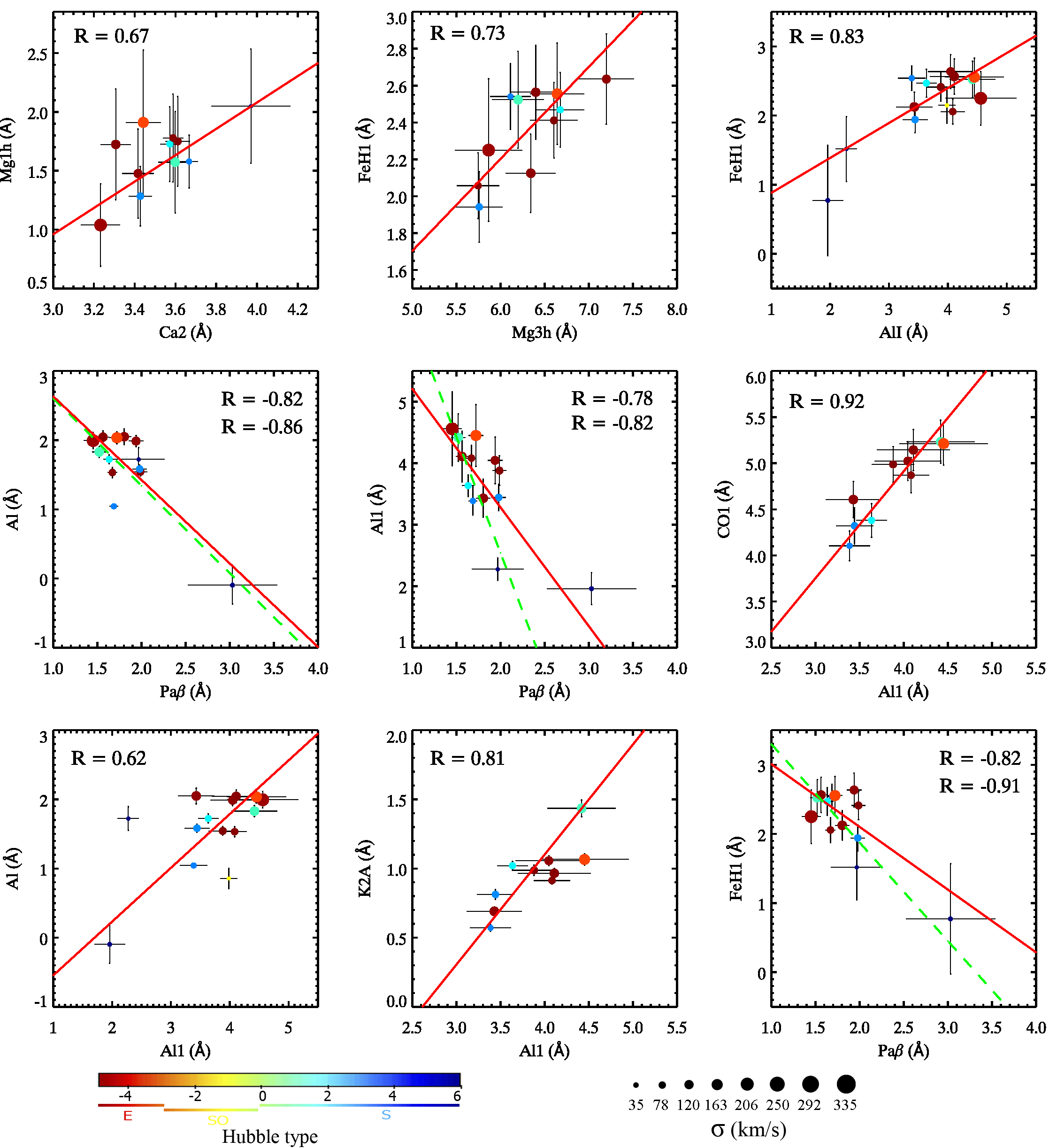}
\caption{The stronger correlations between the NIR line-strength
  indices for the sample galaxies. Galaxies are colour coded according
  to their Hubble type from the Lyon Extragalactic Database
  \citep{makarov2014}. The size of each circle is proportional to the
  velocity dispersion of the galaxy. The best-fitting linear
    relations for all the sample galaxies (red solid line) and
    excluding NGC~584, NGC~636, and NGC~2613 (green dashed line) are
    shown with their Pearson coefficient.}
\label{fig:corr_nir}
\end{figure*}

We found that only 20 out of the possible 780 correlations between the
NIR line-strength indices and velocity dispersion are strong, while
about 100 correlations can be classed as moderate
(Fig. \ref{fig:corr_map_nir}). The strongest correlations between the
NIR line-strength indices are shown in Fig.~\ref{fig:corr_nir} and
include the Mg1h-Ca2 and FeH1-Mg3h correlations already discovered by
\citet{riffel2019}. In these cases, the slope and Pearson correlation
coefficient obtained for our sample galaxies are consistent with those
found by \citet{riffel2019}. The correlations between the Al1 and CO1
indices ($R = 0.92$) and between the Al1 and FeH1 indices ($R = 0.83$)
are strongest ones. The Pa$\beta$ index does not show any strong
correlation with any of the line-strength index based on hydrogen
lines, which are in general very weak. On the other hand, the
Pa$\beta$ index tightly correlates with the Al and FeH1 indices
(Fig. \ref{fig:corr_nir}).

In Fig.~\ref{fig:corr_nir} we also show the distribution of the
morphological types of the sample galaxies. We found that the EWs of
metal line-strength indices (e.g., Al and FeH1) are larger for the
early-type galaxies and smaller for the late-type ones (see also
\citealt{riffel2019}).
   NGC~3423 and NGC~7424 have a distinct behaviour with respect
     the other sample galaxies as they tend to lie in a separate
     region of the parameter space. We suggest this is due to the fact
     that they are the only two late-type, low velocity dispersion,
     young and metal-poor galaxies of our sample (Paper I). Therefore,
     their stellar populations are different from that of the
     early-type galaxies that constitute most of our galaxy
     sample. For this reason, we considered these two galaxies
     interesting to unveil the global trends observed in line-strength
     indices moving from early to late-type galaxies. However, we are
     aware that this issue needs further investigation by increasing
     the number of young late-type galaxies.

\subsection{NIR-$\sigma$ correlations}
\label{sec:nirsigma}

In the optical regime, the correlation between some optical
line-strength indices (e.g., Mg2, \Fem, and H$\beta$) and velocity
dispersion was first found for early-type galaxies
\citep[e.g.,][]{burstein1988, bender1993, bernardi1998, mehlert2003}
and then it was demonstrated to hold also for late-type galaxies too
\citep[e.g.,][]{moorthy2006, sanchez2006, morelli2008}. In the NIR
domain, \citet{cenarro2003} discovered a strong anti-correlation
between the CaT index and velocity dispersion in the central regions
of 35 early-type galaxies. \citet{barroso2003} found that this
anti-correlation also holds for spiral bulges. Finally, the
correlation between $K$-band line-strength indices (e.g., Na, Ca, and
CO) and velocity dispersion was pointed out by \citet{silva2008,
  marmol2009, rock2017}.

\begin{figure*}
\centering
\includegraphics[width=19truecm,angle=0]{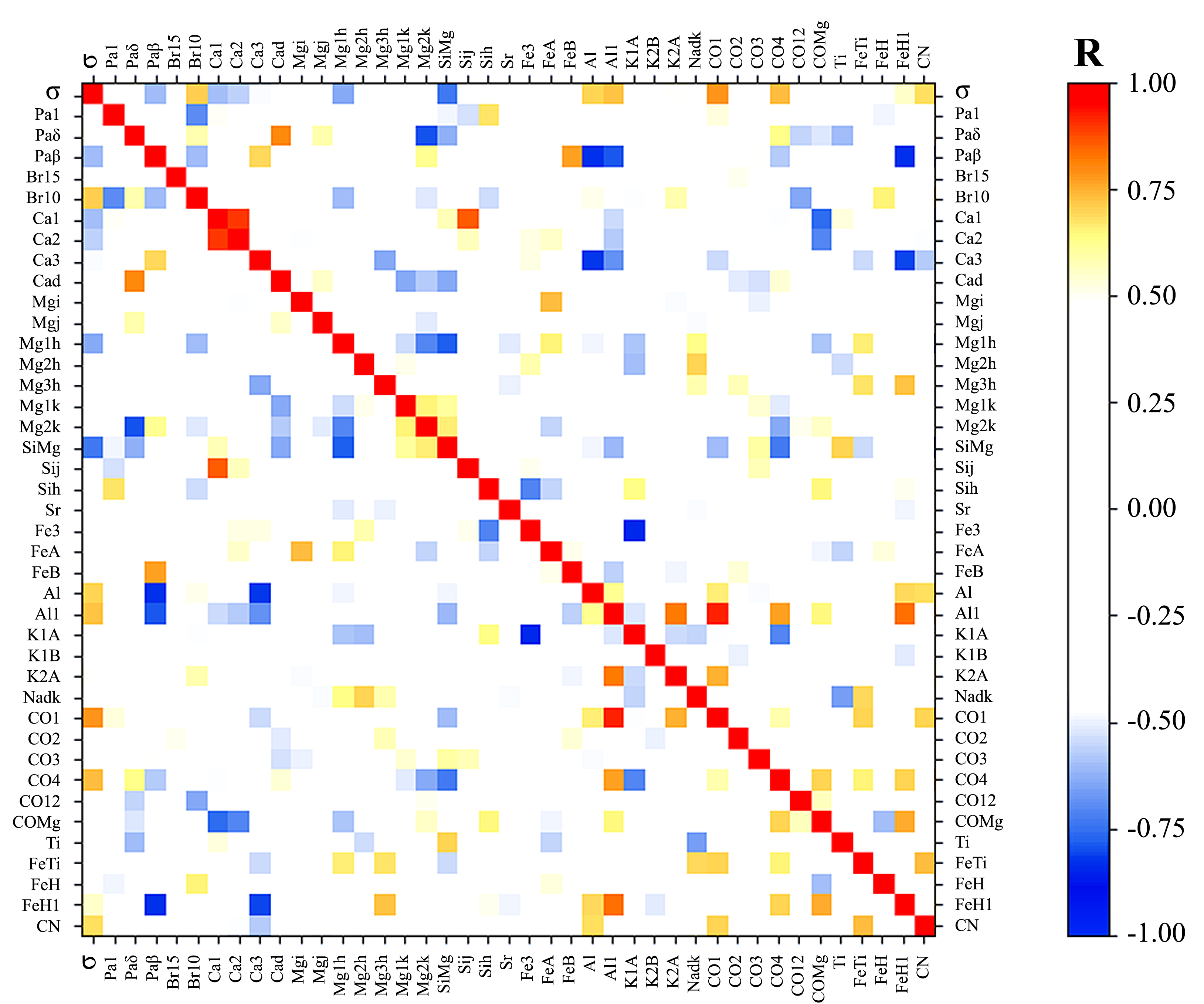}
\caption{The correlation matrix for the final set of 40 NIR line-strength
  indices and velocity dispersion for the sample galaxies.  Different
  colours correspond to different degrees of correlation according to
  Pearson statistics. For sake of clarity, only the moderate and
  strong correlations and anti-correlations (with Pearson coefficient
  $0.5<|R|<0.7$ and $|R|>0.7$, respectively) are shown. The
  correlations of the NIR line-strength indices with themselves and
  velocity dispersion with itself (giving $R=1$ by definition) lie on
  the plot diagonal.}
\label{fig:corr_map_nir}
\end{figure*}

\begin{figure*}
\centering
\includegraphics[width=18truecm]{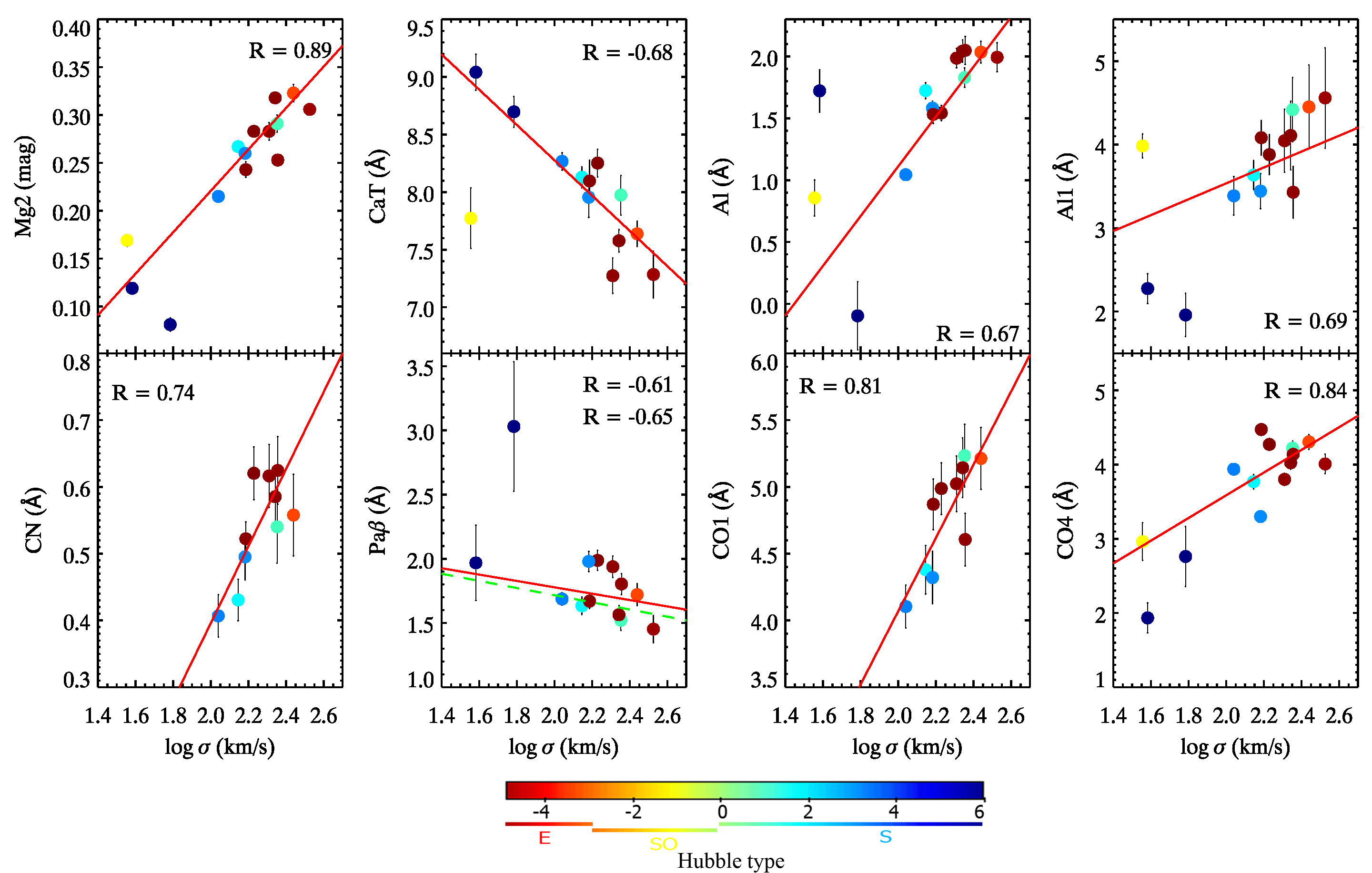}
\caption{The stronger correlations between the NIR line-strength
  indices and velocity dispersion for the sample galaxies. The symbols
  are the same as in Fig.~\ref{fig:corr_nir}.}
\label{fig:corr_sigma}
\end{figure*}

We investigated the correlation between the NIR line-strength indices
and velocity dispersion of the sample galaxies and show their Pearson
coefficient in Fig. \ref{fig:corr_map_nir}. We found that several NIR
line-strength indices (Pa$\beta$, Br10, Ca1, Ca2, Mg1h, SiMg, Al, Al1,
CO1, CO4, FeH1, and CN) show moderate-to-strong correlation with
velocity dispersion. The strongest correlations between the NIR
line-strength indices and velocity dispersion are shown in
Fig. \ref{fig:corr_sigma}.

We found that the Al and Al1 indices are tightly correlated with
velocity dispersion, while the Pa$\beta$ index shows a moderate
anti-correlation with it. This trend resembles what observed for the
H$\beta$ index in the optical regime \citep{ganda2007, morelli2008}.
Among the CO indices, the CO1 and CO4 indices display the stronger
correlation with velocity dispersion. In the first and second panel of
Fig.\ref{fig:corr_sigma} correlations with the velocity dispersion are
shown for the optical Mg2 and the CaT. Their trends are consistent with
previous findings \citep{burstein1988, cenarro2003}.

\section{Age and metallicity indicators}
\label{sec:correlation_optical}

In the previous sections, we found that some indices associated to metal
elements do correlate among themselves (e.g. Al, All, Pa$\beta$, CO1,
FeH1). The linear correlation between these indices, in addition to
the fact that their values are larger in more massive galaxies, as it
happens for the metal indices in the optical regime
\citep{sanchez2006, morelli2008}, is particularly interesting since it
suggests a connection between them and the metallicity of the
galaxies.

In Section \ref{sec:results} we also found a trend between values of
Pa$\beta$ and the morphological type of the galaxies. Even if there
are some notably exceptions, the old spheroids of ellipticals and
early-types show on average lower values of Pa$\beta$ with respect to
the bulges of spirals galaxies (Fig. \ref{fig:corr_nir} and
Fig. \ref{fig:corr_sigma}). Since in general spheroids of ellipticals
and early-type are older than the bulges of late-type galaxies, this
result is a hint that Pa$\beta$ could be a good candidate to trace the
age of the stellar populations.

To better investigate the possible use the NIR line-strength indices
as age and metallicity indicators, we compared them with a set of
optical indices known to be the best diagnostics of unresolved
stellar populations in galaxies. We used the H$\beta$ index as age
indicator, the \Fem, Mg2, Mgb, and the \MgFe\/ \citep{thomas2003} as
metallicity tracers. We measured their values for all the sample
galaxies in Paper I.

\subsection{Trends with age indicators}
\label{sec:trendage}

The H$\beta$ index is widely used as age indicator for unresolved
stellar populations \citep[e.g.,][]{worthey1994, lee2000}. We expect
that some the line-strength indices based on H we studied could
represent the NIR counterpart of the H$\beta$ index. Except for
$Pa\beta$, we did not find any strong trend with the H$\beta$ index of
the Pa1, Pa$\delta$, Br15, and Br10 indices. However, these indices as
we measured in our spectra are too weak and/or contaminated by
residuals of sky subtraction to allow a robust comparisons with the
H$\beta$ index.

\begin{figure}
\includegraphics[width=8cm,angle=0]{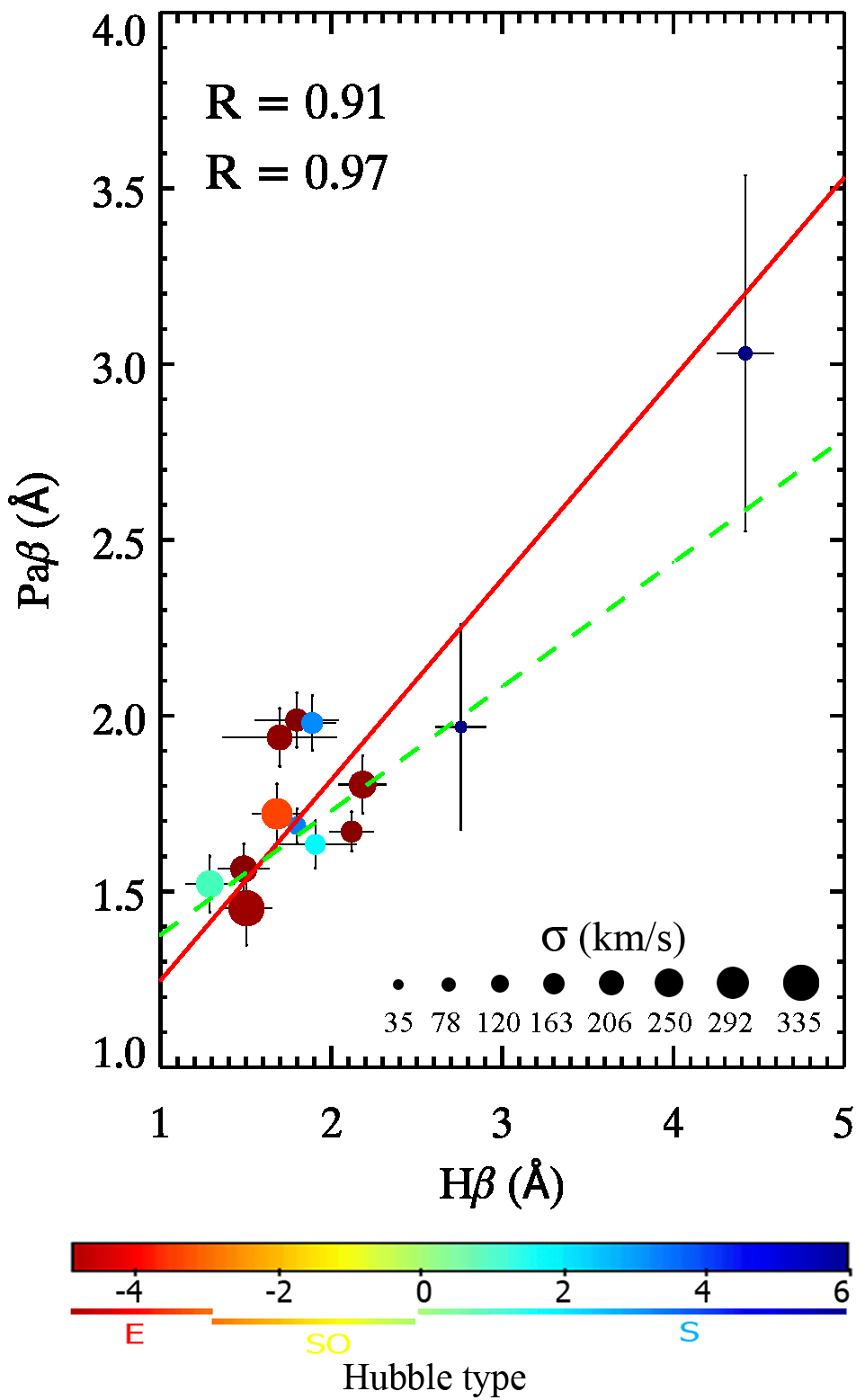}
\caption{The correlation between the NIR Pa$\beta$ index and optical
  H$\beta$ index for the sample galaxies. The symbols are the same as
  in Fig.~\ref{fig:corr_nir}.}
\label{fig:hbeta_corr}
\end{figure}

Fig.~\ref{fig:hbeta_corr} shows the correlation between the H$\beta$
and Pa$\beta$ indices. This correlation is driven by the very young
population resulting from the low SNR spectrum of NGC 7424 ($R=0.91$),
but it still holds ($R = 0.58$) if we do not consider this galaxy.
In Fig.~\ref{fig:hbeta_corr} we noticed that three galaxies seems
  to be shifted with respect to other sample galaxies having a
  $Pa\beta$ index $\sim0.3$ dex larger compared with galaxies of
  similar H$\beta$ index. They are the Sb galaxy NGC~584 and the two
  elliptical galaxies NGC~636 and NGC~2613. They have an
  intermediate-to-old age ($T=6-10$ Gyr) with super-solar metallicity
  ($[Z/{\rm H}]=0.24-0.32$ dex) and $\alpha$/Fe enhancement
  ($[\alpha/{\rm Fe}]=0.20-0.25$ dex). To further investigate their
  behaviour, we fitted a linear relation by excluding the three
  galaxies (Fig. \ref{fig:hbeta_corr}). The Pearson coefficient
  improved from $R=0.91$ to 0.97 and the correlation slope became
  slightly shallower, but it did not change the increasing trend of
  the $Pa\beta$ index with the H$\beta$ index. We expect to strengthen
  this result and better constrain the slope of the $Pa\beta$-H$\beta$
  relation with more young galaxies in the high-end region of the
  H$\beta$ index ($EW=3-5$ \AA).

\subsection{Trends with the metal indices}
\label{sec:corr_metal_lick}

The optical Mg2 and Mgb indices are sensitive to $\alpha$ elements,
while the \Fem\/ index is sensitive to the abundance of the elements of
the iron group. They are used to infer the total metallicity and
$\alpha$/Fe enhancement together with the \MgFe\/ index.

We found that none of line-strength indices based on Fe of our NIR set
shows a strong correlation with the \Fem\/ index, with the partial
exception of two $K$-band FeA and FeB indices. However, considering
the weakness of these lines, we preferred to focus our attention to
the stronger FeH1 index.

\begin{figure*}
\includegraphics[width=18truecm,angle=0]{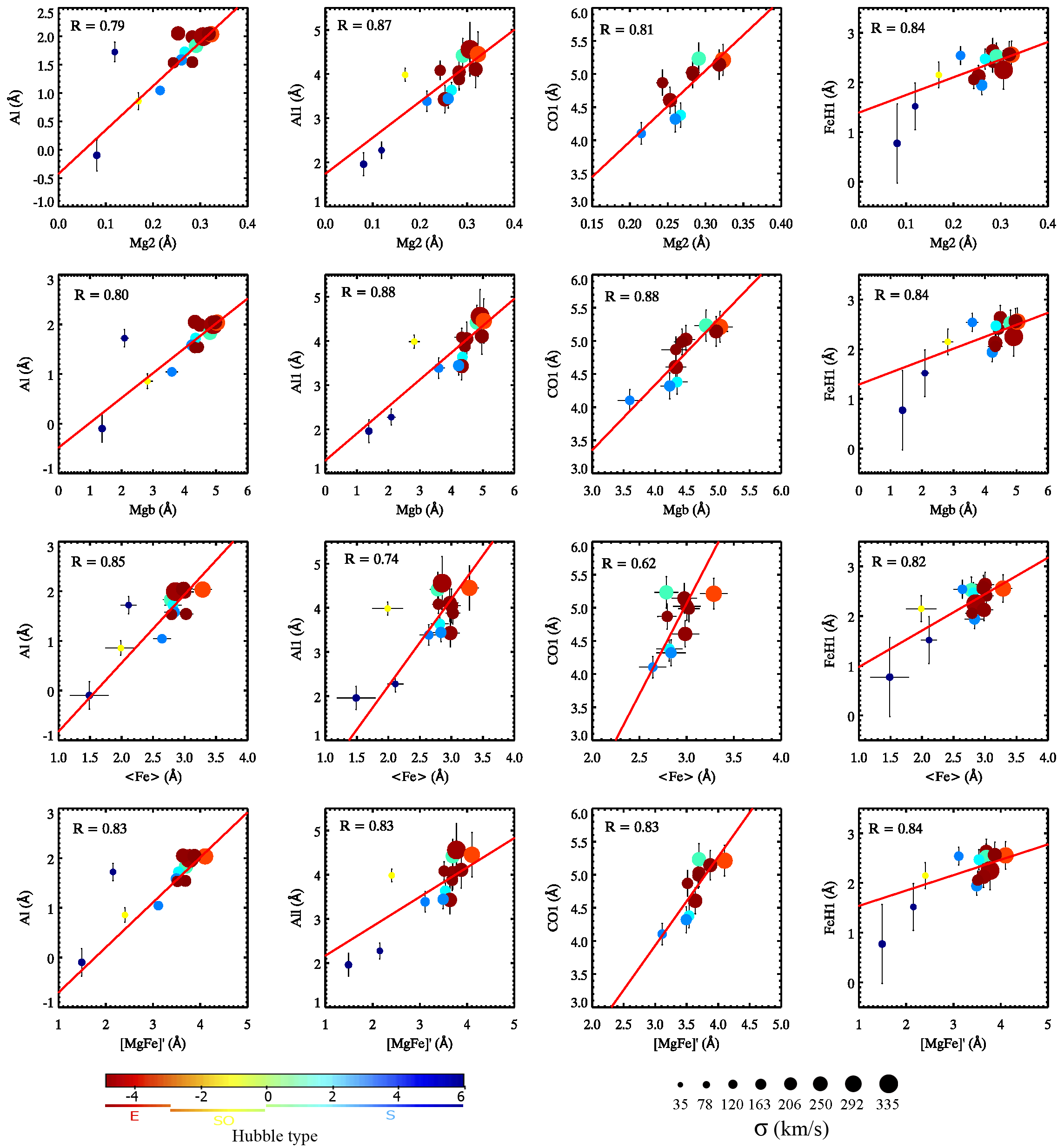}
\caption{The correlations between the NIR Al, Al1, CO1, and FeH1
  indices and optical Mg2, Mgb, \Fem, and \MgFe\/ indices for the
  sample galaxies. The symbols are the same as in
  Fig.~\ref{fig:corr_nir}.}
\label{fig:metal_corr}
\end{figure*}

In Fig. \ref{fig:metal_corr} we present the correlations between the
Al, Al1, CO1, and FeH1 indices and the Mg2, Mgb, \Fem, and \MgFe\/
indices. The strongest correlation between the NIR and optical
line-strength indices are the Al1-Mg2 ($R=0.87$), Al1-Mgb ($R=0.88$),
CO1-Mgb ($R=0.88$), and Al-\Fem\/ ($R=0.85$) correlations. The
correlations between the Al1 and CO1 indices and the \MgFe\/ index are
poorer with respect to those with the Mg2 index.

To improve the sensitivity of the NIR line-strength indices to the
metallicity of the galaxies, we defined two new line-strength indices
by a linear combination of those available in our set. First, we
considered line-strength indices due to same element and for which
Fig.~\ref{fig:corr_map_nir} shows at least a moderate correlation with
some of the optical indices. We did not find any notable improvement with either the Mg or CO indices, whereas we noticed that averaging
the two Al indices into the combined \Alm\/ index as
\begin{equation}
\langle{\rm Al}\rangle = 0.5(Al+Al1)
\end{equation}
improved the correlations with the Mg2, Mgb, \Fem, \MgFe\/ indices. We
further tightened these correlations by including the FeH1 index to
define the following combined [AlFeH] index as
\begin{equation}
  [AlFeH] = Al +0.5Al1 + FeH1
\end{equation}
where we empirically determined the 0.5 coefficient for the Al1 index
to maximise the correlation with the \MgFe\/ index.

\begin{figure*}
\centering
\includegraphics[width=18truecm,angle=0]{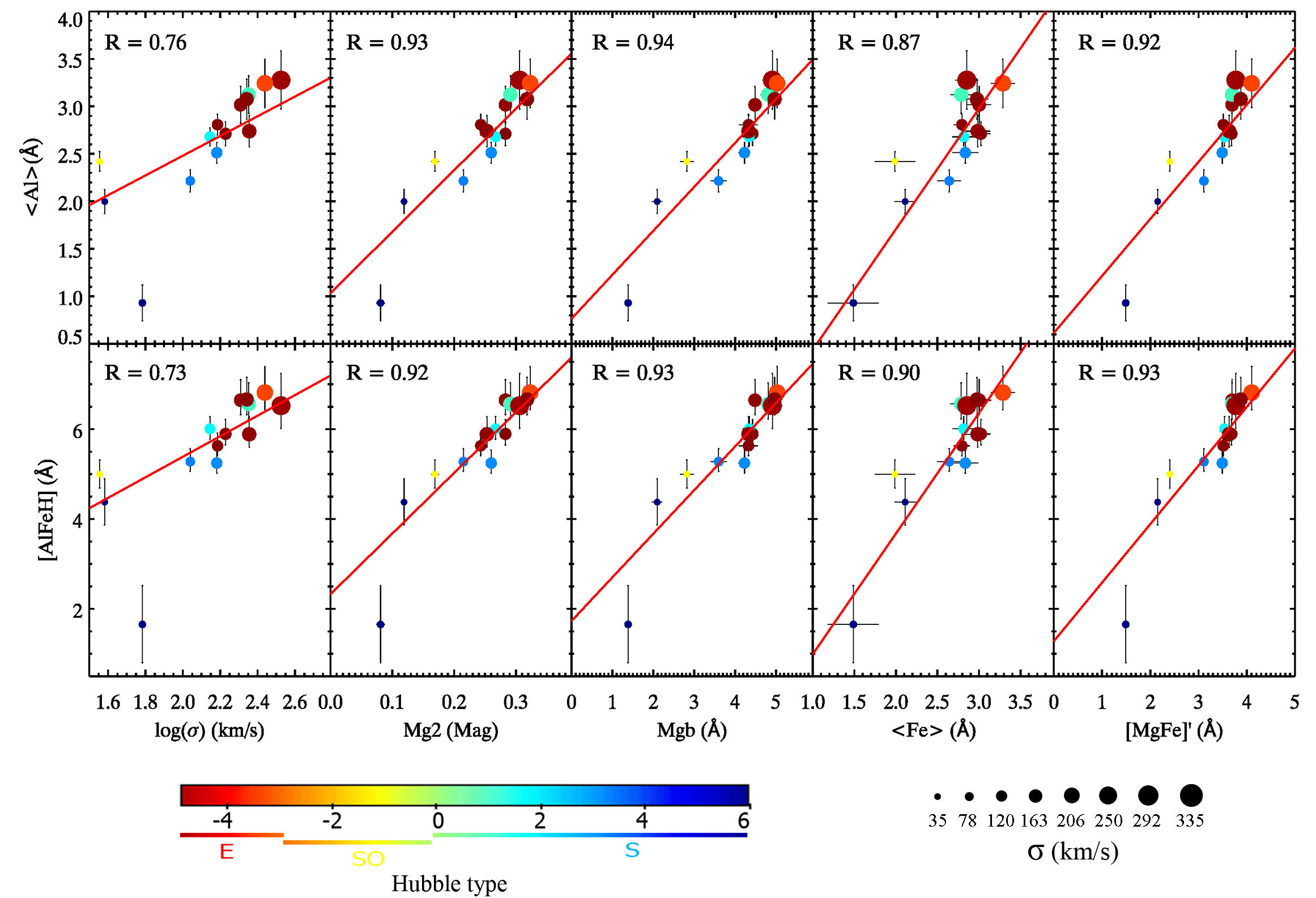}
\caption{The correlations between the NIR \Alm\/ and [AlFeH] NIR indices
  with velocity dispersion and optical Mg2, Mgb, \Fem, and \MgFe\/
  indices for the sample galaxies. The symbols are the same as in
  Fig.~\ref{fig:corr_nir}.}
\label{fig:idx_comb}
\end{figure*}

Fig.~\ref{fig:idx_comb} shows the correlations between the two
composite indices with velocity dispersion and Mg2, Mgb, \Fem, and
\MgFe\/ indices.  These two newly defined line-strength indices are
the most effective NIR indices to correlate with the optical
metallicity indices. The \Alm\/ index traces very well the behaviour of
the Mg indices and maintains a strong correlation with the velocity
dispersion, while the [AlFeH] traces better the behaviour of the \Fem\/
and \MgFe\/ indices.

\begin{figure*}
\centering
\includegraphics[width=9truecm,angle=270]{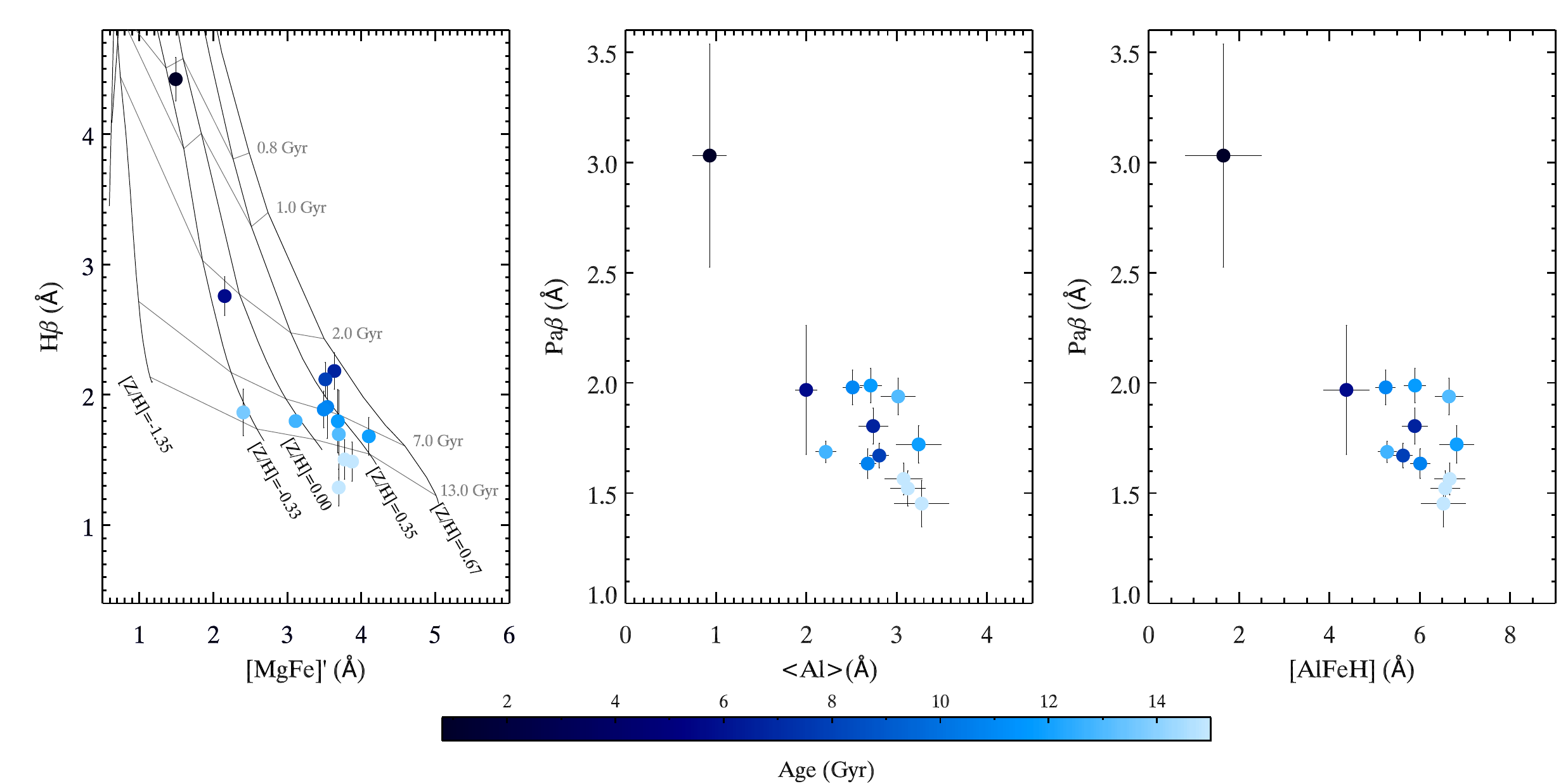}
\caption{The distribution of the measured values of the optical
  H$\beta$ and \MgFe\/ indices (left panel), NIR Pa$\beta$ and \Alm\/
  indices (central panel), and NIR Pa$\beta$ and [AlFeH] indices
  (right panel) for the sample galaxies. The black and grey lines in
  the left panel correspond to the predicted values of the H$\beta$
  and \MgFe\/ indices for a grid of mean ages and total metallicities
  according to the models of \citep{johansson2010} for an $\alpha/Fe$
  enhancement of [$\alpha$/Fe] = 0.3 dex. Galaxies are colour coded
  according to their mean age.}
\label{fig:nir_lick_similarities_age}
\end{figure*}

\begin{figure*}
\centering
\includegraphics[width=9truecm,angle=270]{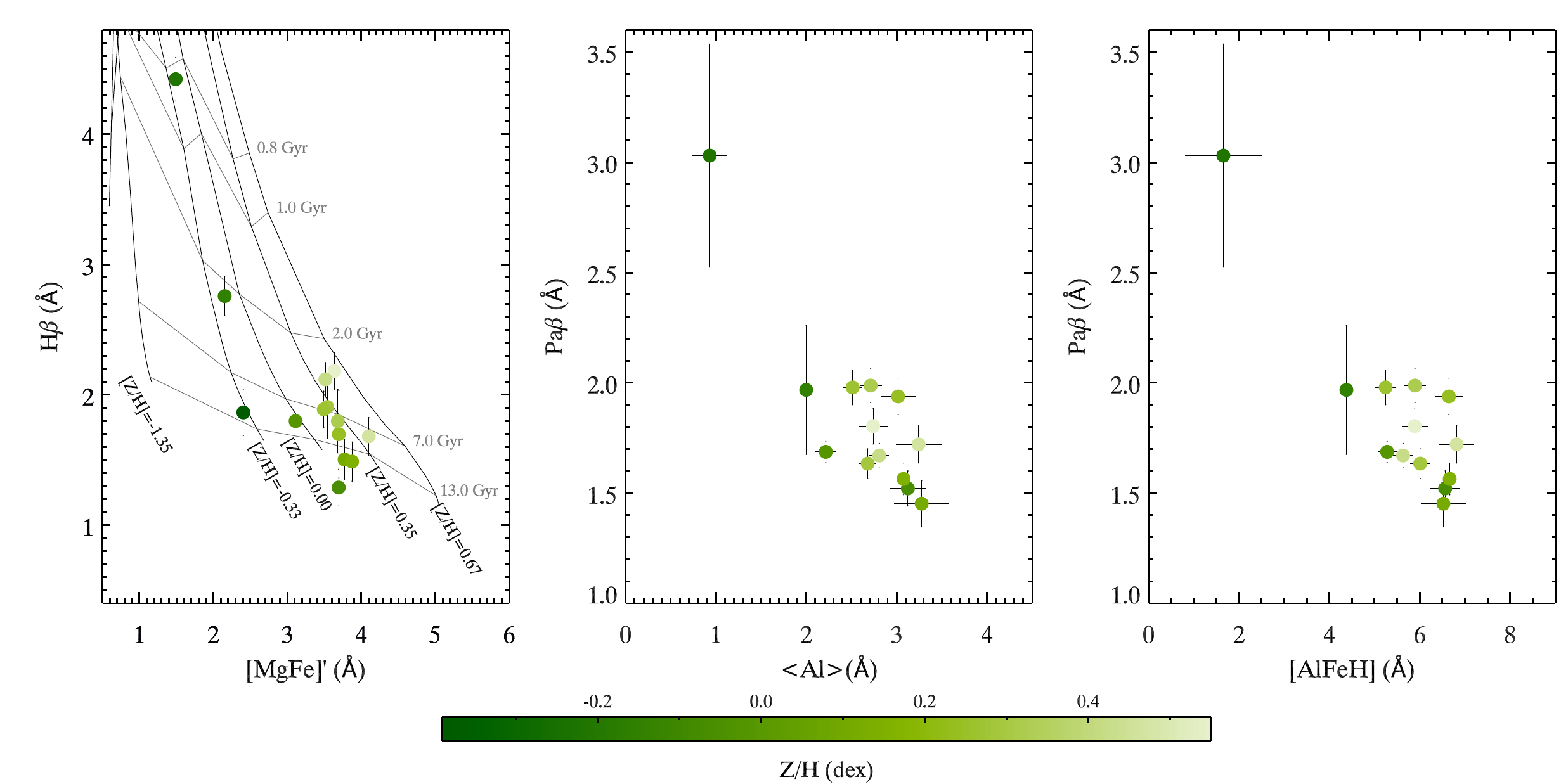}
\caption{The distribution of the measured values of the optical
  H$\beta$ and \MgFe\/ indices (left panel), NIR Pa$\beta$ and \Alm\/
  indices (central panel), and NIR Pa$\beta$ and [AlFeH] indices
  (right panel) for the sample galaxies. The black and grey lines in
  the left panel correspond to the predicted values of the H$\beta$
  and \MgFe\/ indices for a grid of mean ages and total metallicities
  according to the models of \citep{johansson2010} for an $\alpha/Fe$
  enhancement of [$\alpha$/Fe] = 0.3 dex. Galaxies are colour coded
  according to their total metallicity.}
\label{fig:nir_lick_similarities_met}
\end{figure*}

\begin{figure*}
\centering
\includegraphics[width=9truecm,angle=270]{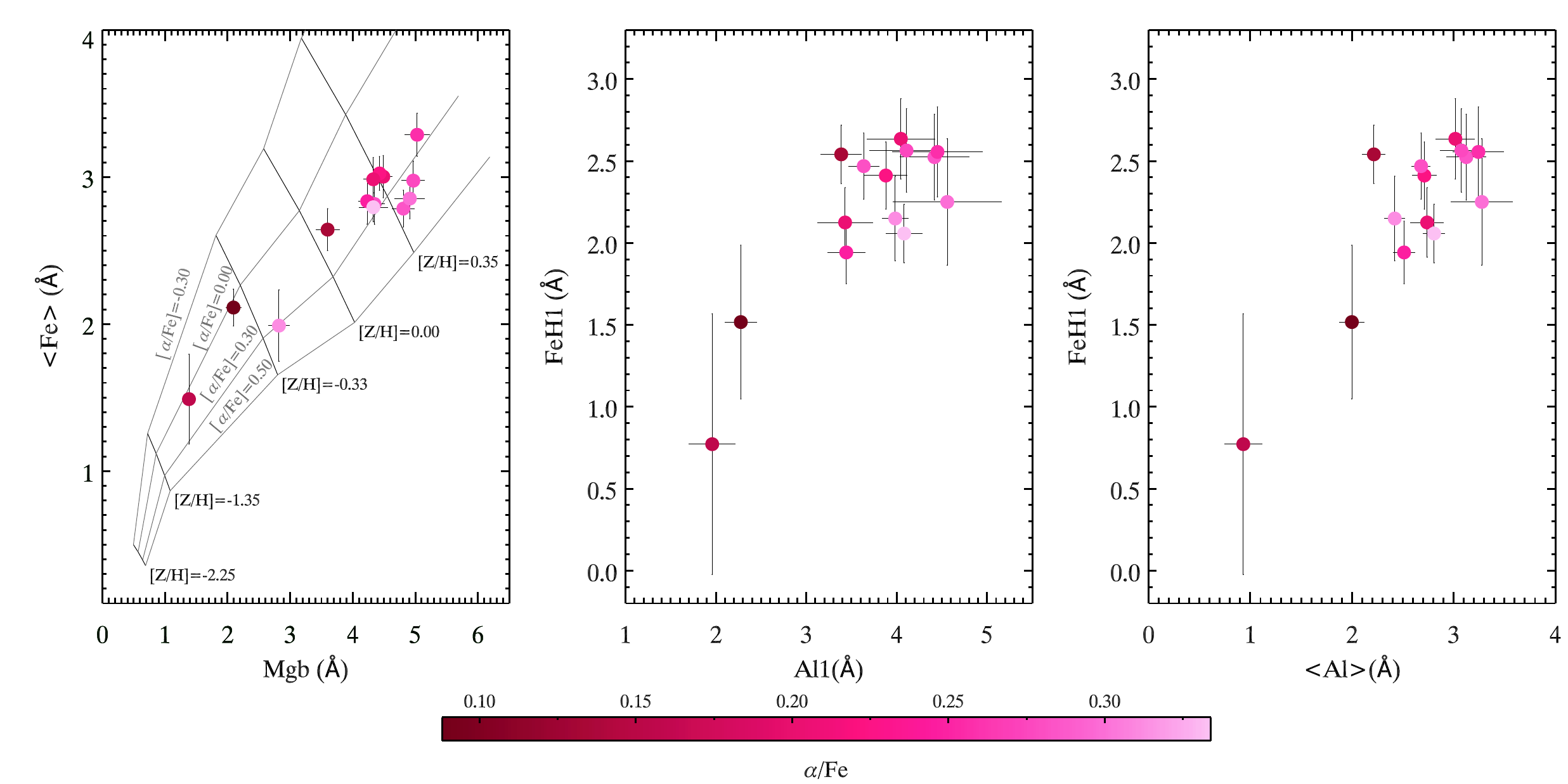}
\caption{The distribution of the measured values of the optical \Fem\/
  and Mgb indices (left panel), NIR FeH1 and Al1 indices (central
  panel), and NIR FeH1 and \Alm\/ indices (right panel) for the sample
  galaxies. The black and grey lines in the left panel correspond to
  the predicted values of the \Fem\/ and Mgb indices for a grid of total
  metallicities and $\alpha$/Fe enhancement according to the models of
  \citep{johansson2010} for a mean age of 7 Gyr. Galaxies are colour
  coded according to their $\alpha$/Fe enhancement.}
\label{fig:nir_lick_similarities_alpha}
\end{figure*}

\section{Discussion}
\label{sec:discussion}

The Al and Al1 indices are among the strongest atomic features after
the Mg indices and they show a behaviour similar to the Mgb and Mg2
indices. The origin of the only stable isotope of $^{27}$Al is still
debated. According to \cite{nordlander2017}, it can be produced in
neutron-rich environments or via proton capture. In galaxies, the main
sources may be the SNII explosion and AGB stars, while SNIa explosion
produce few or null aluminium.  \cite{pignatari2016} draw very similar
conclusions and pointed out the fact that the production condition of
$^{27}$Al is very similar to those of Mg. Considering also the
correlation with the velocity dispersion (Fig. \ref{fig:corr_sigma}),
the Al indices could be related to the formation and evolution of
galaxies in the same way as the Mg2 and Mgb indices.

In the 0.8-2.5 $\mu$m range, there are only three iron features that
might be efficiently used as metallicity diagnostic. The two $K$-band
FeA and FeB indices have already been investigated \citep{silva2008},
but they are weak and located in a spectral region of low SNR in the
spectra of our sample galaxies.  The FeH1 index is located in a
spectral region relatively free from telluric and atmospheric features
and not contaminated by other lines. In our sample galaxies, it is
stronger than the FeH WingFord band. For stars the FeH1 index shows a
dependence from surface gravity and metallicity (see Figs. D5 and D6
in \citealt{morelli2020}).  For galaxies the strong correlations of
the FeH1 index with the optical metal line-strength indices support
the idea it could be a good metallicity tracer.

The Pa$\beta$ line is the strongest hydrogen feature for all the
galaxies in our sample. We identified three galaxies, namely
  NGC~584, NGC~636, NGC~2613, as possible outliers in the
  H$\beta$-Pa$\beta$ relation (Fig.~\ref{fig:hbeta_corr}) and we
  tested their behaviour in the relations between the Pa$\beta$ index
  and Al, Al1, and FeH1 indices (Fig. \ref{fig:corr_nir}) as
  well as between the Pa$\beta$ index and velocity dispersion
  (Fig.~\ref{fig:corr_sigma}).
 Although in this case the three galaxies are difficult to be
 identified as outliers due to their small shift with respect to the
 bulk of the sample galaxies, we excluded them from the linear
 fit. The new best-fitting relations are also shown in
 Figs. \ref{fig:corr_nir} and \ref{fig:corr_sigma}. The slope and
 Pearson coefficient are nearly the same for the Al-Pa$\beta$ and
 $\sigma$-Pa$\beta$ relations, while the slope is steeper for the
 Al1-Pa$\beta$ and FeH1-Pa$\beta$ relations but the trend is the same
 as before.  We conclude that with the actual number of galaxies and
 their distribution in the parameter space we can not firmly conclude
 on the outlier nature of these galaxies.

The Pa$\beta$ index does not suffer any significant contamination by
other elements. \cite{morelli2020} probed that the index behaviour in
the cool IRTF stars is driven by temperature, with no trends with
surface gravity and metallicity. \cite{cleri2020} showed that the
Pa$\beta$ index could adopted as a tracer of the star formation rate
in galaxies, showing a similar (and somewhat even better) behaviour
with respect to the H$\alpha$ and H$\beta$ indicators. This supports
our findings (Fig. \ref{fig:hbeta_corr}) and it suggests that the
Pa$\beta$ index could be also considered as a good age tracer for
unresolved stellar populations in NIR.

We defined the \Alm\/ and [AlFeH] indices to make available stronger
metallicity and $\alpha$/Fe enhancement indicators in the NIR. The \Alm\/ index tightly
correlates with velocity dispersion and optical Mg2 and Mgb indices
(Fig. \ref{fig:idx_comb}) in agreement with the theoretical findings
of \cite{pignatari2016} and \cite{nordlander2017}. The [AlFeH] index
is more sensitive to total metallicity, like the optical \Fem\/ and \MgFe\/
indices. 

Considering the above correlation, we performed the following
speculative analysis.
In the left panels of Figs. \ref{fig:nir_lick_similarities_age} and
\ref{fig:nir_lick_similarities_met}, the values of H$\beta$ and
\MgFe\/ of the sample galaxies (Paper I) are compared to the model
predictions of \citet{johansson2010} for a stellar population with
supersolar $\alpha$/Fe enhancement [$\alpha$/Fe] = 0.3 dex. In
this parameter space, the mean age and total metallicity of the
galaxies are almost insensitive to the variations of
$\alpha$/Fe enhancement. The sample galaxies are colour coded
according to their mean age
(Fig. \ref{fig:nir_lick_similarities_age}) and total metallicity
(Fig. \ref{fig:nir_lick_similarities_met}) which we derived in
Paper I.
For comparison, the values of \Alm\/ and [AlFeH] of the sample galaxies
are shown as function of Pa$\beta$ in the central and right panels of
Figs. \ref{fig:nir_lick_similarities_age} and
\ref{fig:nir_lick_similarities_met}. The distributions of data points in the
\Alm-Pa$\beta$ and [AlFeH]-Pa$\beta$ diagrams nicely match that of the
\MgFe-H$\beta$ diagram. Indeed, the younger and more metal-poor
galaxies lie in the left upper region, while the older and more
metal-rich galaxies are in the right lower region of all the three
diagrams.  Both the \Alm\/ and [AlFeH] indices are very effective in
discerning total metallicity when compared to the \MgFe\/ index. For
intermediate ages galaxies, the Pa$\beta$ index seems slightly less
efficient than H$\beta$ in disentangling mean ages.

In the left panel of Fig. \ref{fig:nir_lick_similarities_alpha}, the
values of Mgb and \Fem\/ of the sample galaxies (Paper I) are compared
to the model predictions of \citet{johansson2010} for a stellar
populations with an intermediate age of 7 Gyr. In this parameter
space, the total metallicity and $\alpha$/Fe enhancement appear to be
almost insensitive to the variations of age. The sample galaxies in
Fig. \ref{fig:nir_lick_similarities_alpha} are colour coded by their
total $\alpha$/Fe enhancement, which we derived using a linear
interpolation between the model points with the iterative procedure
described in \citet{morelli2008} and in Paper I. The values of
$\alpha$/Fe enhancement for the sample galaxies are reported in
Table~\ref{tab:idx_lick}.
For comparison, the values of Al1 and \Alm\/ of the sample galaxies are
shown as function of FeH1 in the central and right panel of
Fig.~\ref{fig:nir_lick_similarities_alpha}, respectively.  The
distribution of data points in the Al1-FeH1 and \Alm-FeH1 diagrams
nicely match that of the Mgb-\Fem\/ diagram. This is a promising result,
although the absence of galaxies with extremely high/low values of
[$\alpha$/Fe] in our sample makes difficult to evaluate the
effectiveness of these NIR line-strength indices in constraining the
$\alpha$/Fe enhancement.

Following these results, we propose the NIR line-strength indices
Pa$\beta$, \Alm, Al1, FeH1, and as possible counterparts of the
optical indices H$\beta$, \MgFe, Mgb, and \Fem\/ to investigate the
unresolved stellar populations in the NIR domain for spectra with
medium resolution ($R \sim 5000$) and high SNR ($> 100$ \AA$^{-1}$).

\section{Conclusions}
\label{sec:conclusions}

We investigated a set 40 out of the 75 line-strength indices proposed
by \citet{cesetti2013} and \citet{morelli2020} in the $I$, $Y$, $J$,
$H$, and $K$ bands for a sample of 14 nearby galaxies observed with
the ESO XShooter spectrograph. The galaxies span all the Hubble
morphological sequence with a mean age range $0.8 \leq {\rm age} \leq
15 Gyr$ and a total metallicity range $-0.39\leq [Z/{\rm H} \leq 0.55$
  dex. Up to date, this is the largest set of line-strength indices
  measured and tested in the NIR domain.

We found that some of the studied NIR line-strength indices are
promising candidate to constrain the properties of unresolved stellar
populations in galaxies. To further explore this idea, we compared
them with the most-widely used optical age and metallicity indicators.

The Al, Al1, CO1, and FeH1 indices were found to be strongly
correlated with the optical Mg2 and Mgb indices sensitive to
$\alpha$/Fe enhancement and with the \Fem\/ and \MgFe\ sensitive to
total metallicity. The Pa$\beta$ index is tightly correlated with the
H$\beta$ index sensitive to mean age.

We defined two new combined indices \Alm\/ and [AlFeH] to build stronger
metallicity and $\alpha$/Fe enhancement indicators in the NIR. The
\Alm\/ index tightly correlates with velocity dispersion and optical Mg2
and Mgb indices, while the [AlFeH] index is more correlated with the
\Fem\/ and \MgFe\/ indices.

For our sample galaxies, we found a similar distribution of the data
points in the optical \MgFe-H$\beta$ age-metallicity diagnostic
diagram and in our two NIR counterparts given by the \Alm-Pa$\beta$
and [AlFeH]-Pa$\beta$ diagrams. We also found a similar distribution
of the data points in the optical $\alpha$/Fe enhancement-metallicity
diagnostic diagram and in our two NIR counterparts given Al1-FeH1 and
\Alm-FeH1 diagrams. This means that a these new sets of NIR
line-strength indices can be taken as a promising starting point to
derive the mean age, total metallicity, and total $\alpha$/Fe
enhancement in unresolved galaxies.

Our next step of our work will be extending the galaxy sample to the
young ages ($<5$ Gyr) and to a broader range of metallicities. This
will allow us to better address the differences in stellar populations
of early-type galaxies and spiral bulges. This analysis can also be
useful as a benchmark to test the new generation of NIR SSP models
domain \citep[e.g.][]{conroy2012, rock2015, vazdekis2016, rock2017}
which still require to be fine-tuned with observational data
\citep{riffel2019}.

\begin{table}
\centering
\caption{The values of $\alpha$/Fe enhancement for the sample galaxies
  estimated with the models of \citet{johansson2010}.}
\label{tab:idx_lick}

\begin{tabular}{cc}
\hline
\hline

Galaxy &  [$\alpha$/Fe] (dex) \\
\hline
NGC~584	        &	0.20	$\pm$	0.07	\\
NGC~636	        &	0.22	$\pm$	0.04	\\
NGC~897	        &	0.28	$\pm$	0.11	\\
NGC~1357	&	0.28	$\pm$	0.04	\\
NGC~1425	&	0.13	$\pm$	0.04	\\
NGC~1600	&	0.30	$\pm$	0.08	\\
NGC~1700	&	0.30	$\pm$	0.05	\\
NGC~2613	&	0.25	$\pm$	0.06	\\
NGC~3115	&	0.26	$\pm$	0.02	\\
NGC~3377	&	0.33	$\pm$	0.07	\\
NGC~3379	&	0.28	$\pm$	0.11	\\
NGC~3423	&	0.08	$\pm$	0.08	\\
NGC~4415	&	0.31	$\pm$	0.10	\\
NGC~7424	&	0.15	$\pm$	0.06	\\

\hline
\end{tabular}
\end{table}

\section*{Acknowledgements}

EMC, EDB, and AP are supported by MIUR grant PRIN 2017
20173ML3WW$\_$001 and Padua University grants DOR1885254/18,
DOR1935272/19, and DOR2013080/20. LC acknowledges financial support
from Comunidad de Madrid under Atracci\'on de Talento grant
2018-T2/TIC-11612 and from Spanish Ministerio de Ciencia, Innovaci\'on
y Universidades through grant PGC2018-093499-B-I00.

%%%%%%%%%%%%%%%%%%%%%%%%%%%%%%%%%%%%%%%%%%%%%%%%%%
\section*{Data Availability}
The reduced sample galaxies spectra used in this paper have been presented and analysed in \citep{francois2019}. The reduced spectra are available upon request to the authors.

%%%%%%%%%%%%%%%%%%%% REFERENCES %%%%%%%%%%%%%%%%%%

% The best way to enter references is to use BibTeX:
\clearpage
\bibliographystyle{mnras}
\bibliography{bib}

% Don't change these lines
\bsp	% typesetting comment
\label{lastpage}
\end{document}